\documentclass{article}       
\usepackage{epsfig}
\usepackage{natbib}
\usepackage{amssymb}
\usepackage{amsmath}
\usepackage{color}
\usepackage{graphicx}
\begin{document}

\title{Velocity derivatives in a high Reynolds number turbulent boundary layer.  Part III: Optimization of an SPIV experiment for derivative moments assessment
}


\author{Jean Marc Foucaut$^1$  \and  William K. George$^2$ \and Michel Stanislas$^3$  \and Christophe Cuvier$^1$}
\maketitle
\noindent
$^1$ Univ. Lille, CNRS, ONERA, Arts et Metiers Institute of Technology, Centrale Lille,
UMR 9014 - LMFL - Laboratoire de M\'{e}canique des Fluides de Lille - Kamp\'{e} de F\'{e}riet,
F-59000 Lille, France\\
$^2$ Visiting Pr., Centrale Lille, F59651 Villeneuve d'Ascq, France\\
$^3$ Pr. Emeritus, Centrale Lille, F59651 Villeneuve d'Ascq, France
\medskip



\date{Received: date / Accepted: date}

\maketitle

\begin{abstract}
An SPIV experiment using two orthogonal planes simultaneously was performed in the LML boundary layer facility to specifically  measure all of the derivative moments needed to estimate the dissipation rate of the Turbulence Kinetic Energy. The Reynolds number was $Re_\theta = 7500$ or $Re_\tau = 2300$. A detailed analysis of the errors in derivative measurements was carried out, as well as applying and using consistency checks derived from the continuity equation. The random noise error was quantified, and used to ``de-noise'' the derivative moments. A comparison with a DNS channel flow at comparable Reynolds number demonstrated the capability of the technique. The results were further validated using the recent theory developed by \cite{georgestanislas20}.  The resulting data have been extensively used in parts I and II of the present contribution to study near wall dissipation (\cite{stanislas20,george20}). An important result of the present work is the provision of reliable rules for an accurate assessment of the dissipation in future PIV experiments.
\end{abstract}

\section{Introduction}
\label{sec:1}

Most experiments in wall turbulence using PIV seek understanding of the flow organization. This objective generally involves also gathering information from the velocity fields. The basic statistics extracted from the velocity (mean and RMS velocities, Reynolds stresses...) are of prime interest when looking at the momentum equation, but they do not provide much information as soon as the energy transfer is of interest. For this, derivatives of the velocity fields have to be computed to access parameters based on velocity gradients such as the rate-of-strain tensor and its invariants, vorticity or the dissipation rate of Turbulence Kinetic Energy (both as a tensor in the Reynolds stress transport equation or as a scalar in the Turbulence Kinetic Energy equation). The dissipation rate, say $\varepsilon$, is a key parameter of the turbulence and it needs to be modeled to properly represent turbulence with the RANS equations. It is also a key to understanding the behaviour of the smallest scales of turbulence. In part I (\cite{stanislas20}) and part II (\cite{george20}) of this contribution, the results of an SPIV experiment were analysed in detail together with those of a DNS at comparable Reynolds number to obtain a better understanding of the turbulence dissipation. In the present paper, attention will be focussed on the experimental approach used to measure the fluctuating velocity gradients and to obtain the results discussed in those two previous papers. 

\subsection{The basic equations \label{basiceqns}}

All the relevant turbulence equations are presented and discussed in detail in parts I and II, so only the most essential will be repeated here.   Most important, of course, is the scalar dissipation rate which appears in the transport equation for the Turbulence Kinetic Energy given by:
 
\begin{equation}
\varepsilon = \nu \left[\left\langle \frac{\partial u_i}{\partial x_j}\frac{\partial u_i}{\partial x_j} \right\rangle + \left\langle \frac{\partial u_i}{\partial x_j}\frac{\partial u_j}{\partial x_i} \right\rangle\right]
\label{eq:01}
\end{equation}
This is of course just twice the viscosity times the mean square fluctuating strain rate, $\varepsilon =2 \nu \langle s_{ij} s_{ij} \rangle$. To compute this full dissipation it is necessary to measure all the components of the instantaneous gradient tensor and to compute the variance of each term together with a few covariances. This corresponds to a total of twelve terms which can be organized in three groups:

\begin{eqnarray}
\varepsilon &=& \nu \left\{2\left[\left\langle \frac{\partial u_1}{\partial x_1}\frac{\partial u_1}{\partial x_1} \right\rangle + \left\langle \frac{\partial u_2}{\partial x_2}\frac{\partial u_2}{\partial x_2} \right\rangle + \left\langle \frac{\partial u_3}{\partial x_3}\frac{\partial u_3}{\partial x_3} \right\rangle \right]\right.   \nonumber \\ 
&+&\left[\left\langle \frac{\partial u_1}{\partial x_2}\frac{\partial u_1}{\partial x_2} \right\rangle  + \left\langle \frac{\partial u_2}{\partial x_1}\frac{\partial u_2}{\partial x_1} \right\rangle + \left\langle \frac{\partial u_1}{\partial x_3}\frac{\partial u_1}{\partial x_3} \right\rangle \right.  \hspace{16mm}\nonumber \\
&+&\left. \left\langle\frac{\partial u_3}{\partial x_1}\frac{\partial u_3}{\partial x_1} \right\rangle + \left\langle \frac{\partial u_2}{\partial x_3}\frac{\partial u_2}{\partial x_3} \right\rangle+\left\langle \frac{\partial u_3}{\partial x_2}\frac{\partial u_3}{\partial x_2} \right\rangle \right] \nonumber \\  
&+& 2 \left[\left\langle \frac{\partial u_1}{\partial x_2}\frac{\partial u_2}{\partial x_1} \right\rangle + \left\langle \frac{\partial u_1}{\partial x_3}\frac{\partial u_3}{\partial x_1} + \left\langle \frac{\partial u_1}{\partial x_3}\frac{\partial u_3}{\partial x_1} \right\rangle\right] 
\right\}
\label{eq:02}
\end{eqnarray}
All of these are difficult to measure accurately. The dissipation rate $\varepsilon$ as given by equation (\ref{eq:01}) combines both derivatives and statistical computations. The tensorial counterpart of the dissipation, say $\varepsilon_{ik}$, is even more complicated and is included in Part II. Because of these difficulties, the turbulence community over the past 90 years has made several different hypotheses to simplify these measurements. Three of these hypotheses ``local isotropy'', ``local axisymmetry'' and ``local homogeneity'') were examined in detail in part II of this contribution (\cite{george20}).

In the present contribution the objective is to review the methods used to determine the dissipation rate $\varepsilon$ of the TKE and all the components of the dissipation rate tensor $\varepsilon_{ij}$. And also to describe  the dedicated Stereoscopic PIV experiment carried out in a turbulent boundary layer at high Reynolds number.

\subsection{Brief review of SPIV \label{briefSPIV}}

PIV is an imaging technique, based on a physical principle completely different from Hot Wire Anemometry (Heat transfer) or Laser Doppler Velocimetry (Doppler shift). It is very dependent on the characteristics of the CCD or CMOS sensor (which have made tremendous progress in the last twenty years) and on the properties of the laser light. It is well-known to evidence a high level of noise, which in turbulence is a problem at high spatial frequency where the signal is quite low. This is a real drawback when trying to measure velocity derivatives representative of the dissipative scales. Stereoscopic PIV is now an established method to characterize turbulent flow. It has been mostly used to look at the instantaneous structures in turbulence, but many researchers also use it to compute statistics such as mean velocity, Reynolds stress tensor, probability density function or spectra (\cite{adrian00,foucaut11,herpin11}). SPIV allows the measurement of the three components of the velocity in a plane with an accuracy at best of 1-2\% (0.1 pixel) if the experiment is properly designed. For a turbulent flow, it opens the unique capability of studying the organization of the turbulence. Generally, due to the limited spatial resolution of  PIV, the smallest scales of the flow cannot be investigated. One difficulty in the data processing comes from the noise amplification when derivatives have to be computed~(\cite{foucaut02}). However, the velocity gradients are necessary to determine the vorticity and shear. They are also essential for the detection of vortices when using the $Q$ criterion or the swirling strength, both useful tools to study the flow organization.

The possibilities of the SPIV technique can be significantly enlarged by the use of light polarization. Based on this laser light property, it is possible to record velocity fields in two crossed-planes simultaneously.
When the planes are parallel, this method is called dual plane stereoscopic PIV ~(\cite{kahler00}).
The dual plane technique allows the measurement of two velocity fields with an adjustable time delay or spatial separation between them. By varying this delay the space-time correlation of the velocity field can be computed. By varying the separation, the 3D spatial correlation can be obtained. \cite{ganapathisubramani05a} used the dual plane technique to get the full gradient tensor and to study the near wall flow structures. If both planes are perpendicular instead of parallel, this can provide information about the spatial properties of the flow (e.g.,~ \cite{hambleton06},~\cite{ganapathisubramani06}) as a first step toward full 3D3C information as provided for example by TomoPIV or Holographic PIV.

\subsection{The goals of this paper \label{sec-goals}}

In the present paper, Stereoscopic PIV in two crossing-planes is used to compute all the derivatives of the three velocity components in a turbulent boundary layer. The main limitation of PIV is that it only resolves a range of scales which is between the size of the field of view for the largest scales and that of the interrogation window for the smallest ones (c.f., \cite{foucaut04}). Due to these spatial resolution limits and to the PIV noise, access to the small scales relevant to dissipation is not an easy task. Emphasis is consequently put particularly here on the derivative filter choice and on the measurement noise management. The primary interest of the present approach is that everything possible was done to try to minimise the noise contribution, first on the velocity fields themselves, second on the derivatives and finally on the derivative moments which are necessary to access the dissipation rate of Turbulence Kinetic Energy (TKE). In the following these different steps will be detailed and, as much as possible quantified, with the aim of providing guidelines for future measurement of the dissipation with Stereoscopiv PIV. Finally the results will be compared to a recently developed theory by \cite{georgestanislas20} predicting the PIV noise. These results  provide a better understanding of the noise sources and how to minimize them.

\section{Experimental setup}
\label{sec:2}

Various aspects of the experiment and methodology were of necessity included in Parts I and II.  They will described in much more detail here to fill in the gaps for the experimentalists.

\subsection{The wind-tunnel }

\begin{figure*}[ht]
	\resizebox{1.\linewidth}{!}{\includegraphics[scale=1]{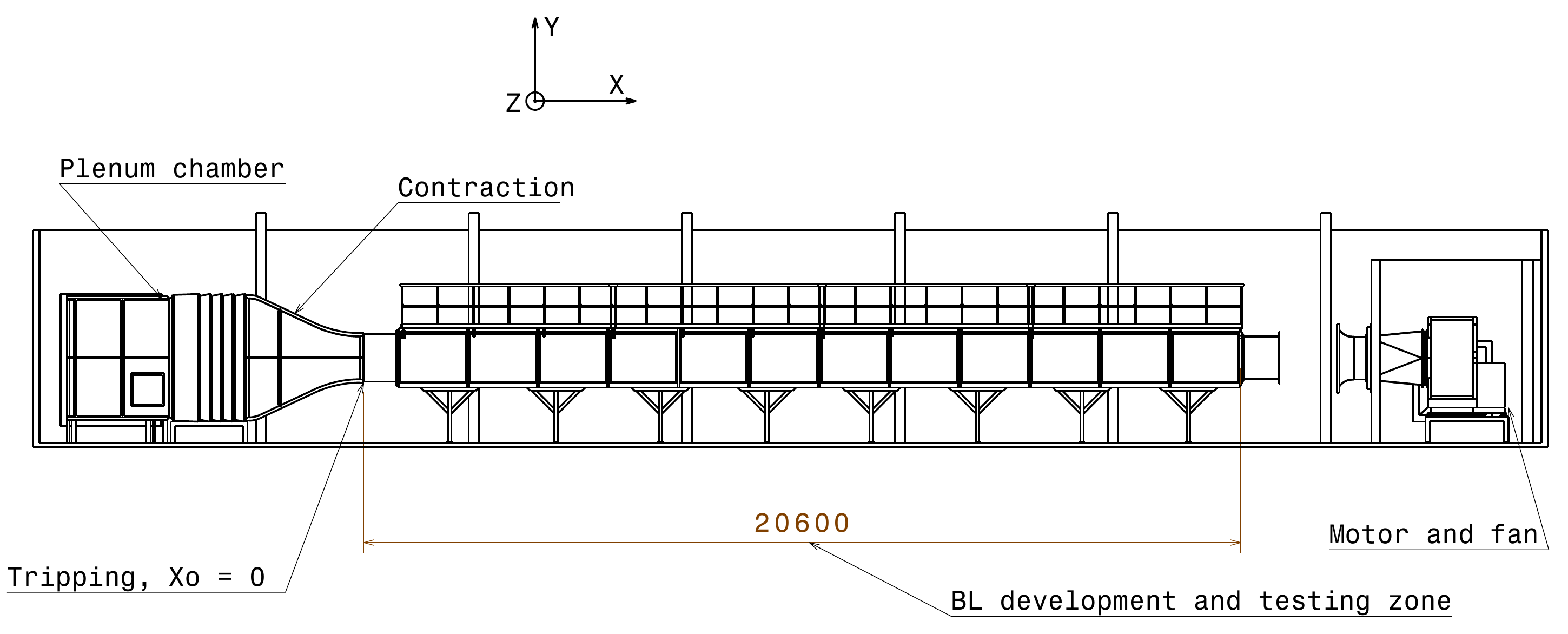}}
	\caption{Side view of the LMFL boundary layer wind  tunnel}
	\label{fig:1}      
\end{figure*}

 The experiment was carried out in the LMFL turbulent boundary layer wind tunnel sketched in figure~\ref{fig:1}. The test section is 1 m high, 2 m wide and 20 m long to allow the development of the boundary layer. The boundary layer thickness can reach 30 cm at the end of the test section. A slight favourable pressure gradient exist and has been characterized by \cite{carlier05}. It is of the order of 1 Pa/m. Two important characteristics of this wind tunnel are that first it is temperature-regulated at better than 0.5C, which allows accurate Hot Wire measurements with very long samples; second that it is fully transparent on 4 sides and down to the wall, which allows a lot of flexibility in the PIV and SPIV set-ups. The present experiment was carried out at a Reynolds number $R_\theta = 7500$ ($Re_\tau = 2300$) which corresponds to a free stream velocity of 3  m/s. 

\subsection{The optical setup, seeding and processing}
\label{setup}

\begin{figure*}[ht]
	\resizebox{0.85\linewidth}{!}{\includegraphics[scale=1]{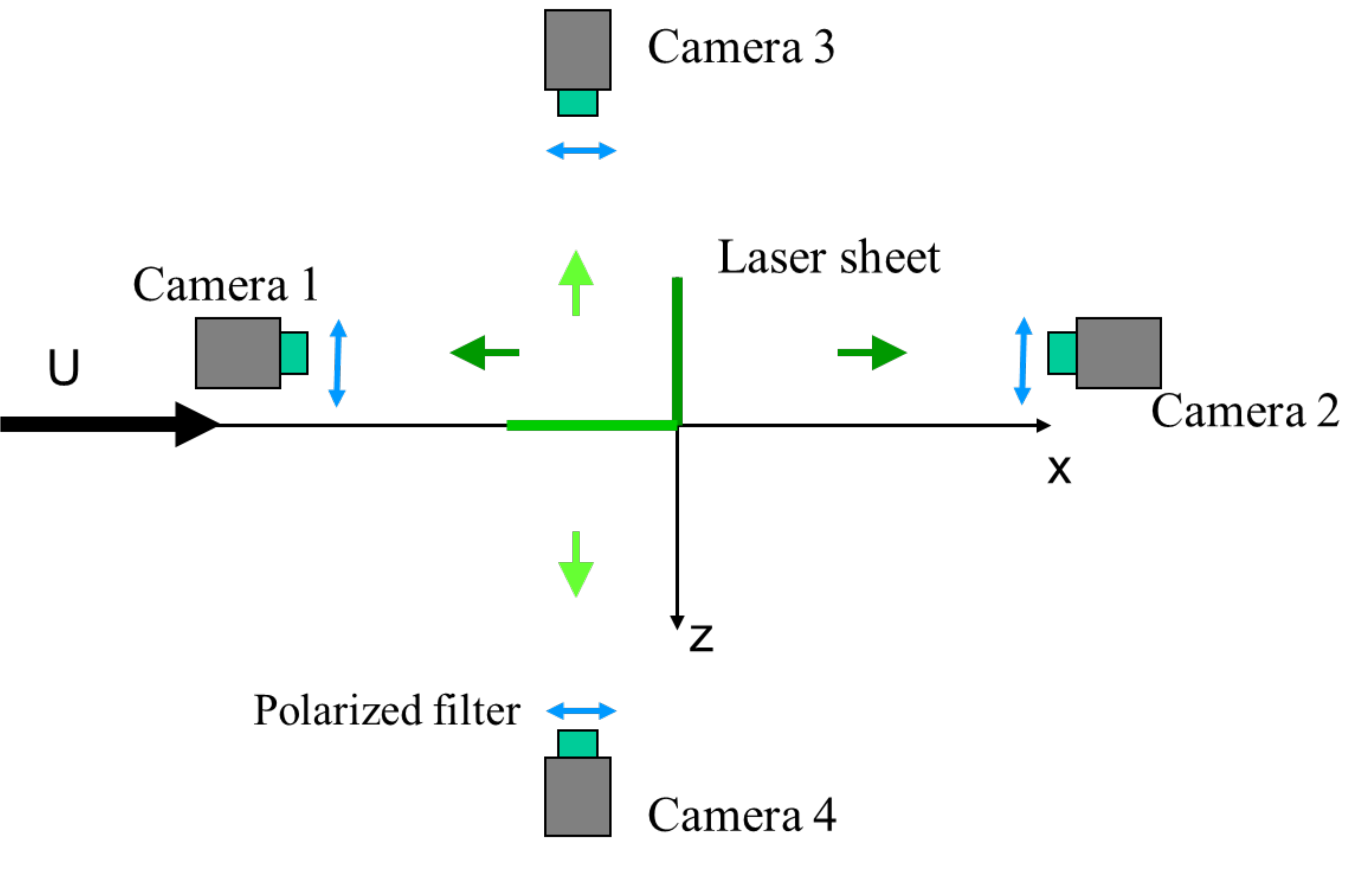}}
	\caption{Top view of the double SPIV experimental set-up}
	\label{fig:2}      
\end{figure*}

Figure~\ref{fig:2} shows a top-view of the SPIV set-up used for this experiment. The laser light sheets were generated with the help of a BMI YAG laser system specifically developed for LMFL. It is composed of 4 laser cavities which can deliver 200 mJ each. Two cavities are polarized vertically and two horizontally. Both sets of cavities are recombined two-by-two to provide two double pulse light beams orthogonally polarized. Lenses and mirrors generate light sheets slightly larger than the field of view to improve the uniformity of the illumination to introduce them in the wind tunnel. The tunnel walls are made of high quality glass, antireflection coated to minimize spurious light. The two light sheets were introduced through the ceiling of the tunnel and exited through the transparent floor where they where caught in a light trap. 

The field-of-view was imaged with two Stereoscopic PIV systems in two orthogonal crossing-planes, both normal to the wall. One plane was streamwise along ($x_1,x_2$). The second one was spanwise along ($x_3,x_2$), where $x_2$ is the wall normal coordinate. Each system was based on Hamamatsu 2k x 2k pixels cameras,  micro Nikkor 105 mm lenses at f\# 8 and  polarizing filters. The field-of-view of each system is about 12 x 8 $cm^2$ (corresponding to about $640 $ x $ 960$ wall units). Both systems were adjusted so that the two PIV planes share a common wall normal line. Each SPIV system was adjusted to fulfill the Scheimpflug conditions (\cite{willert97}). The seeding was done with Poly-Ethylene Glycol micron particles. 
The particle size was of the order of 1 $\mu m$. In this configuration, the Airy disk diameter is of the order of 12.3 $\mu m$. On the Hamamtsu cameras, this gives a particle image size of the order of 1.8 pixel (c.f. \cite{adrian97}). This was checked on the real images. On average, the particle image size appears to be effectively a bit less than 2 pixel. Following \cite{raffel98}, such a size should minimize the peak-locking effect. 

The images from both cameras were processed with a standard multi-grid and multi-pass algorithm with image deformation (\cite{scarano02}). The analysis was made by the classical FFT-based cross-correlation method with symmetrical integer shift of both windows. Before the final fourth pass, image deformation was applied according to the displacements estimated from the three previous passes to improve the accuracy~(\cite{stanislas07}). A 1-D Gaussian peak fitting algorithm was used for the sub-pixel displacement determination. The final interrogation window size was 24 x 32 pixels for each plane. It was chosen based on the HW turbulence spectrum in order to optimize the signal to noise ratio, following the method proposed by \cite{foucaut04}. Such a size corresponds to a square window in the physical space of 1.4 x 1.4 $mm^2$ (11.6 x 11.6 wall units or 6 x 6 Kolmogorov units). An overlap of 66\% was chosen in order to maximize the spatial resolution while bringing 34\% of new information in each interrogation window. 

\begin{figure*}[ht]
	\resizebox{0.85\linewidth}{!}{\includegraphics[scale=1]{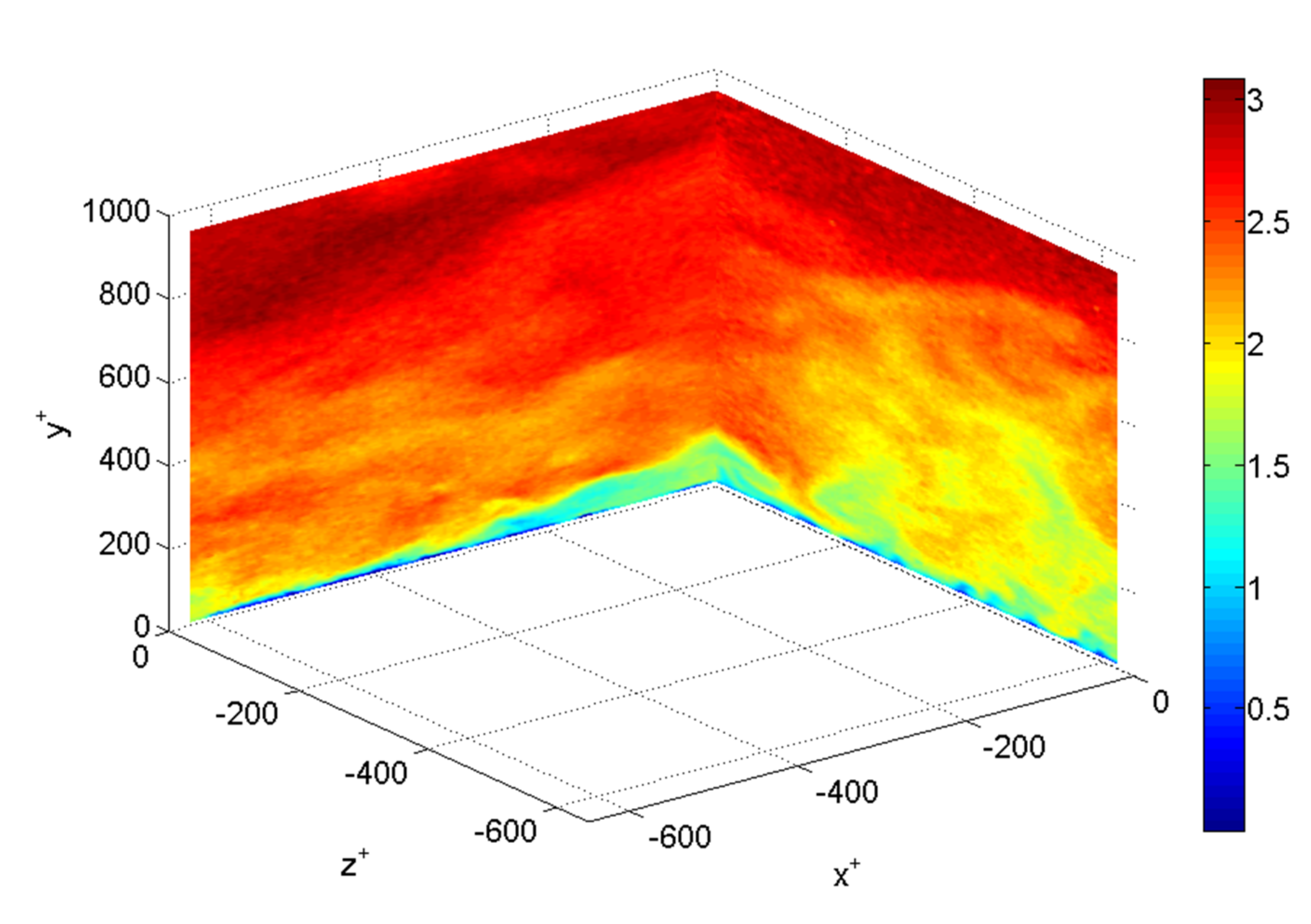}}
	\caption{Example of instantaneous streamwise component velocity fields.}
	\label{fig:3}      
\end{figure*}

The Soloff method using 7 calibration planes was used to reconstruct the three velocity components in the plane of measurement (\cite{soloff97}).  The calibration was done with different targets using crosses. From the set of recorded calibration planes and SPIV images, the misalignment between the light sheet and the calibration plane was corrected (\cite{coudert01}) with a self-calibration method similar to the one proposed by \cite{wieneke05}. The analyses were performed using the MatPIV software adapted and extended by LMFL.  A total of 12000 velocity fields were recorded simultaneously with the two systems and analysed. Figure \ref{fig:3} gives an example of the instantaneous streamwise velocity component $u_1$ in the two planes.

\subsection{Basic flow characteristics}
\label{baseflow}

\begin{figure*}[ht]
	\centering 
	\resizebox{0.9\linewidth}{!}{\includegraphics[scale=1]{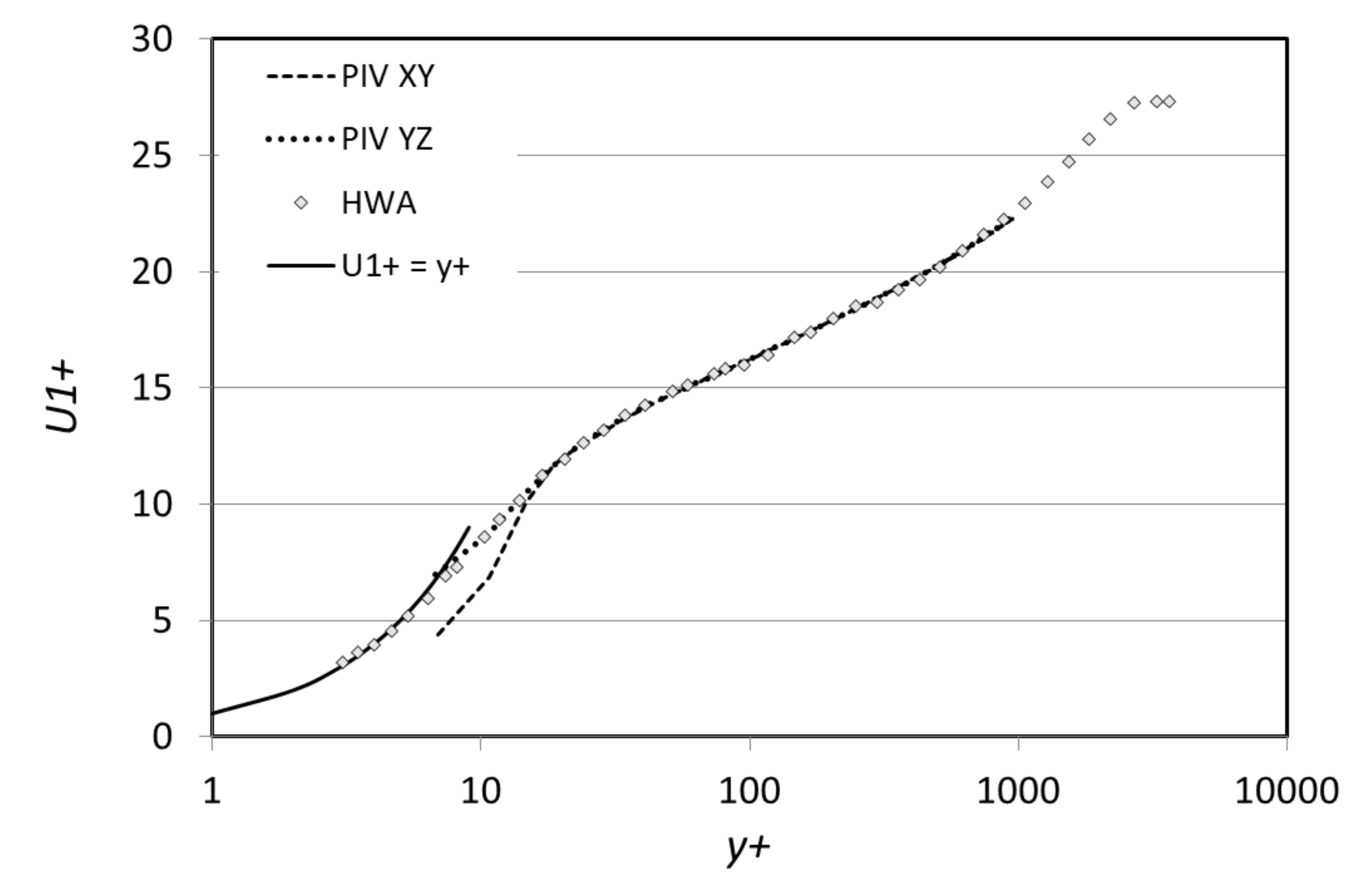}}
	\caption{Mean velocity profiles from the two cross-planes compared to hot-wire data.}
	\label{fig:4}      
\end{figure*}

\begin{figure*}[ht]
	\centering 
	\resizebox{0.9\linewidth}{!}{\includegraphics[scale=1]{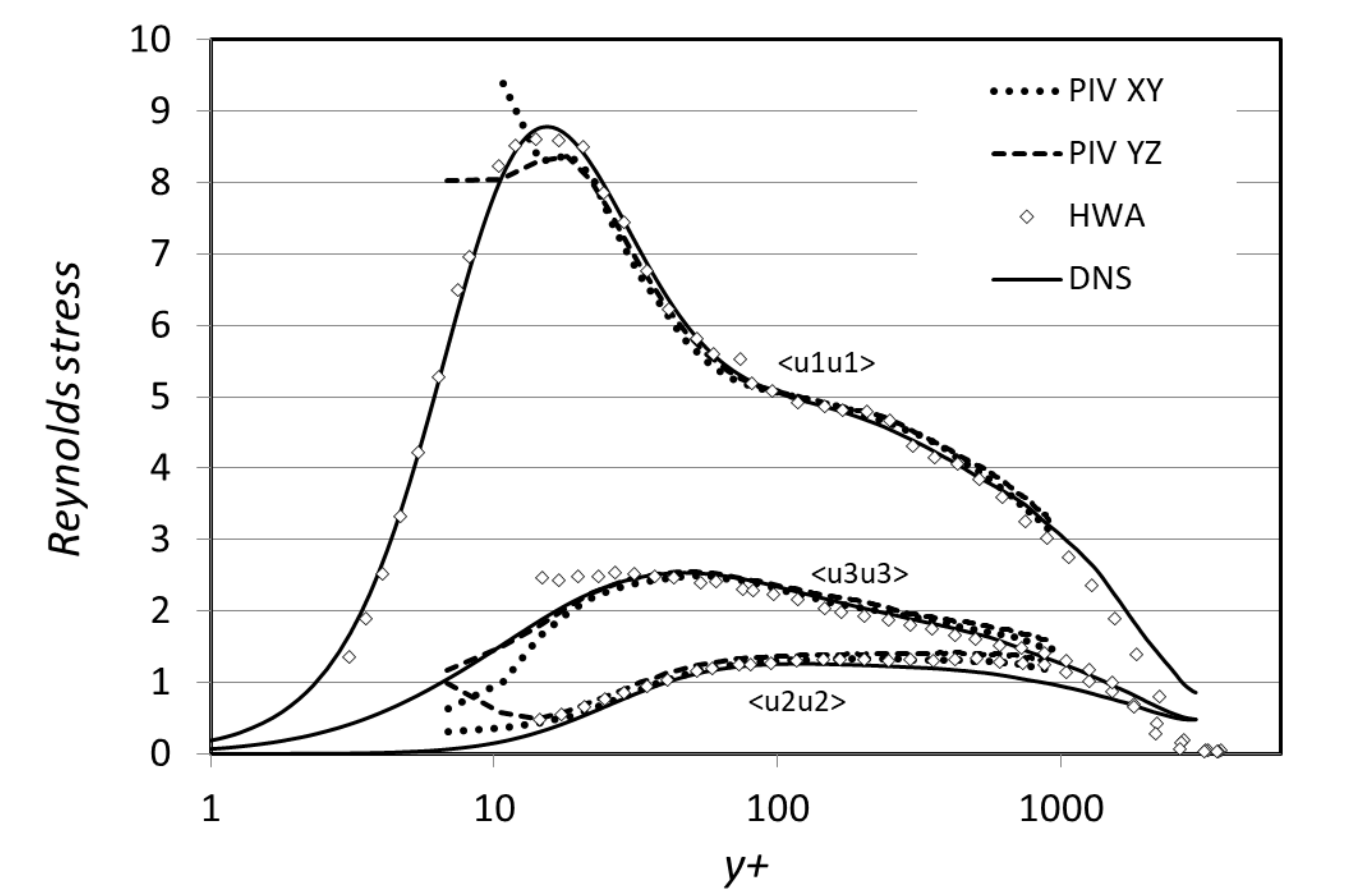}}
	\caption{Turbulent intensity profiles from the two cross-planes compared to hot-wire data.}
	\label{fig:5}      
\end{figure*}

Figures \ref{fig:4} and \ref{fig:5} show the mean streamwise velocity and the turbulent intensity profiles from the two planes respectively, along with HWA results. All these results are in good agreement above 15 wall units, except for the spanwise intensity component in figure \ref{fig:5} which presents a difference below 40 wall units. This discrepancy is most probably due to the mean velocity gradient at the scale of the X-wire probe used, which cannot be taken into account during the calibration. 

\begin{figure*}[ht]
	\resizebox{0.30\linewidth}{!}{\includegraphics[scale=1]{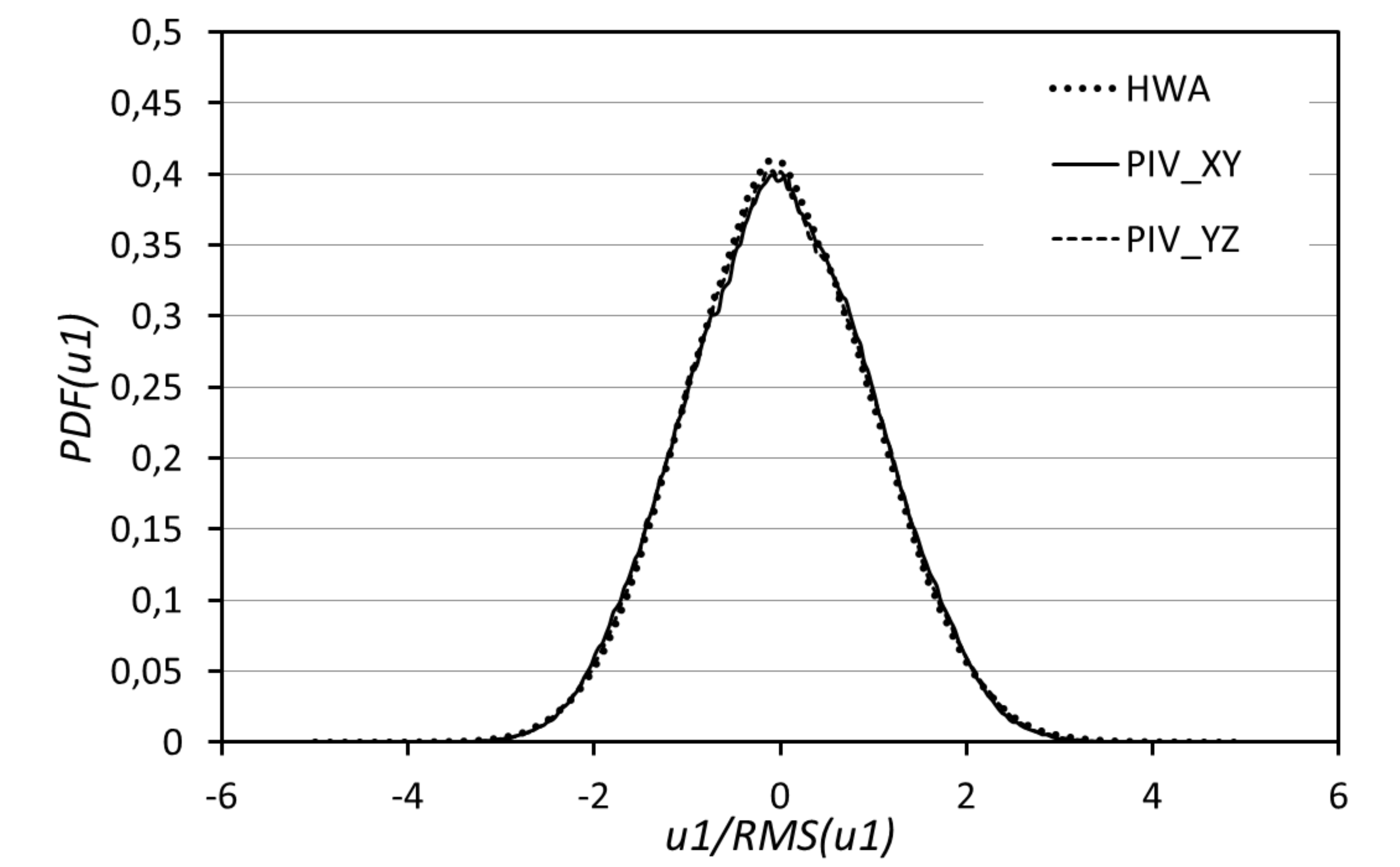}}
	\resizebox{0.30\linewidth}{!}{\includegraphics[scale=1]{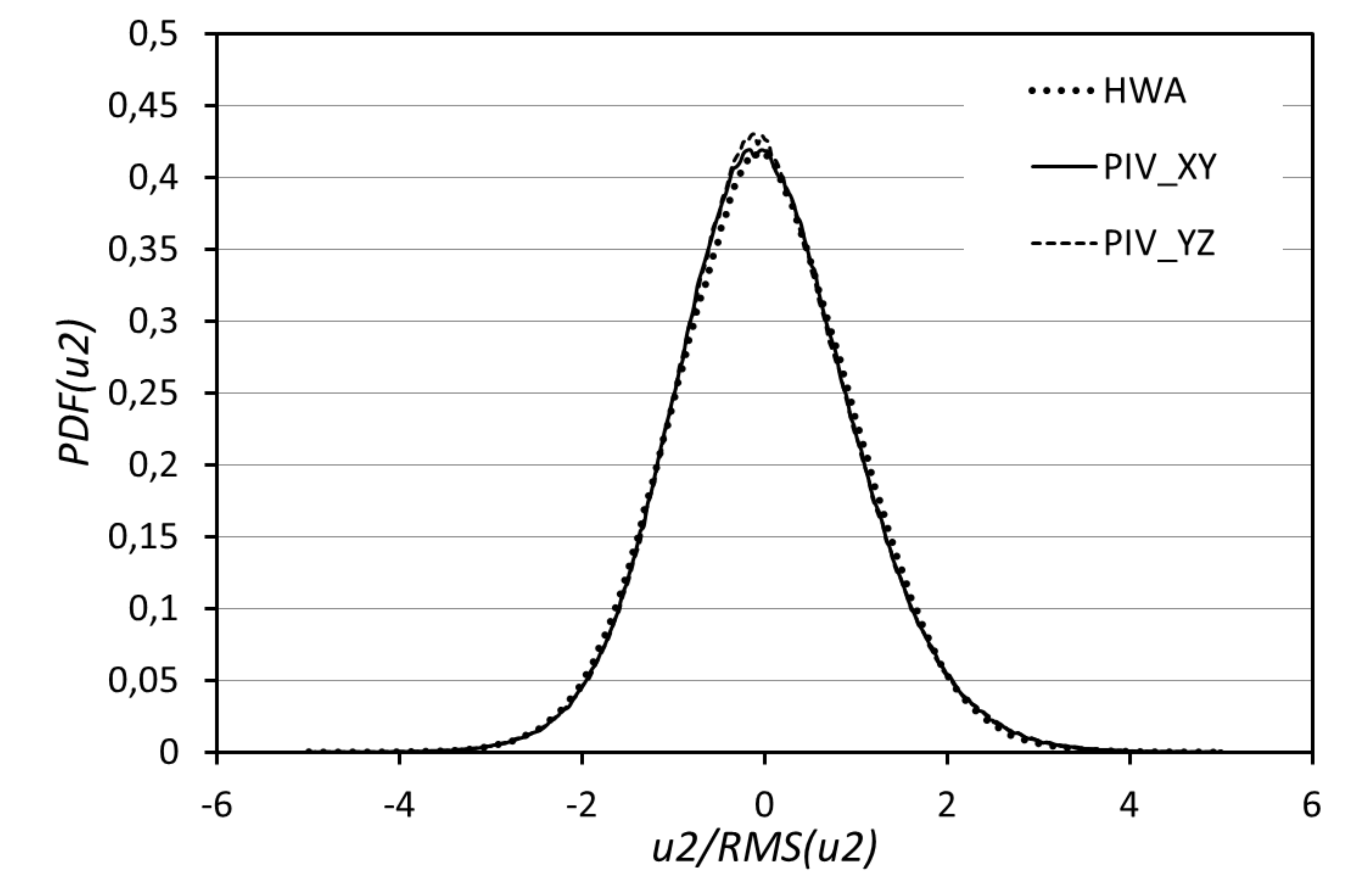}}
	\resizebox{0.30\linewidth}{!}{\includegraphics[scale=1]{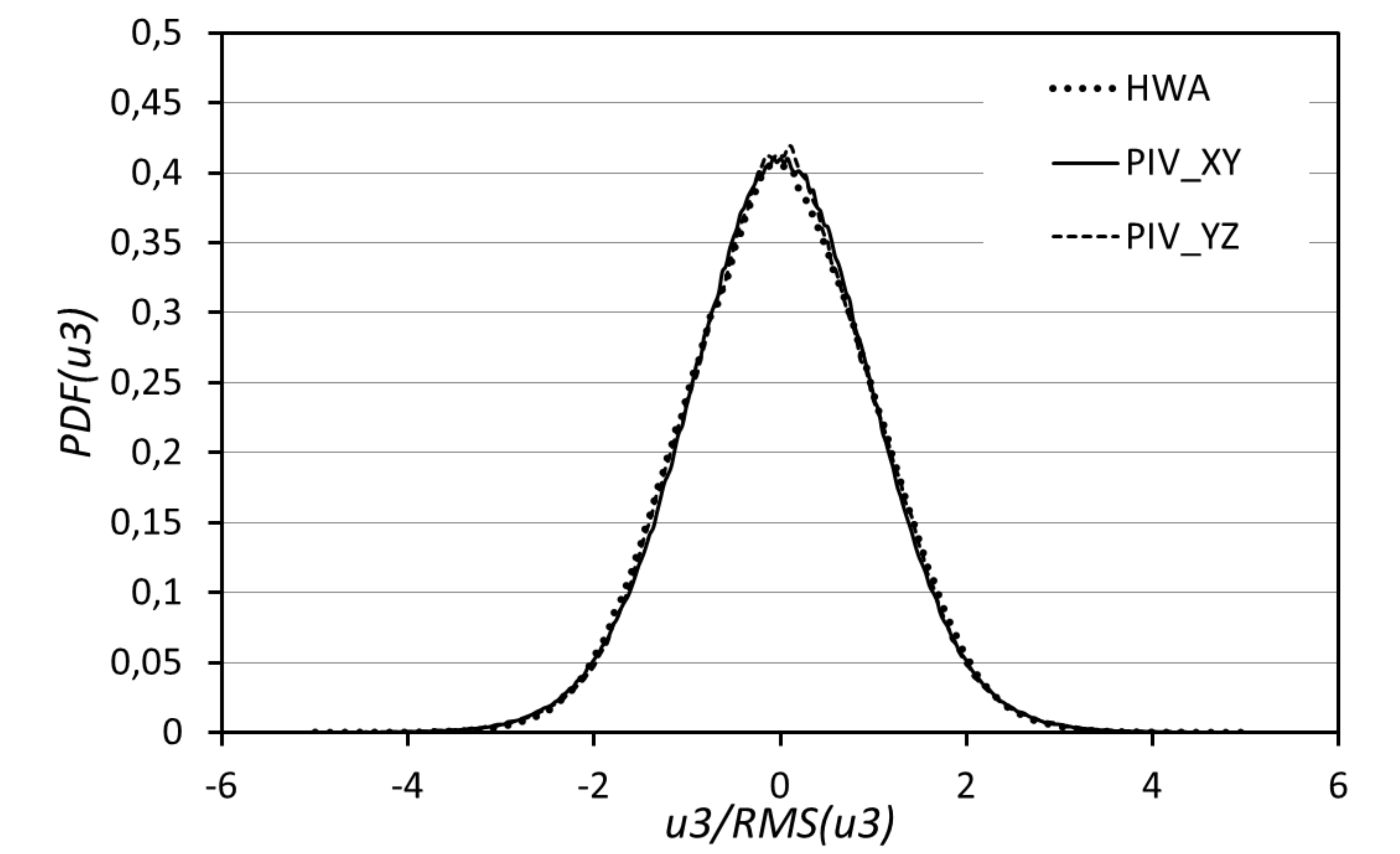}}
	\caption{Comparison of the PDF of the 3 fluctuating velocity components from SPIV and HWA at $y^+ = 100$.}
	\label{fig:6}      
\end{figure*}

Figure \ref{fig:6} shows the probability density functions  of the three fluctuating velocity components from both planes at 250 wall units and compared with hot wire anemometry. A very good agreement is observed between the two SPIV systems and the hot wire results. Also they show no clipping of the velocity due to a saturation of the dynamics  in the plane normal to the flow and no peak locking is in evidence.

\section{Velocity spectra and noise}

Looking now at the spectrum, two important characteristics of the PIV technique have to be looked at: the limited field of view and the fact that an interrogation window of finite size is used. The influence of the IW size will be characterized in the next section when looking at the small scales. 

\subsection{The turbulence spectra}

Note that here the spectra are being computed using spatial transforms of the entire field-of-view, and not by applying Taylor's hypothesis to a time-varying signal at one location. The finite field-of-view multiplies the turbulent velocity field by a rectangular window which cuts off the large scales and generates leakage down the spectrum toward the small scales.  In Fourier spectral space this manifests itself as a ``sinc-squared''-function convolved with the spectrum. 

\begin{figure*}[ht]
	\centering 
	\resizebox{0.9\linewidth}{!}{\includegraphics[scale=1]{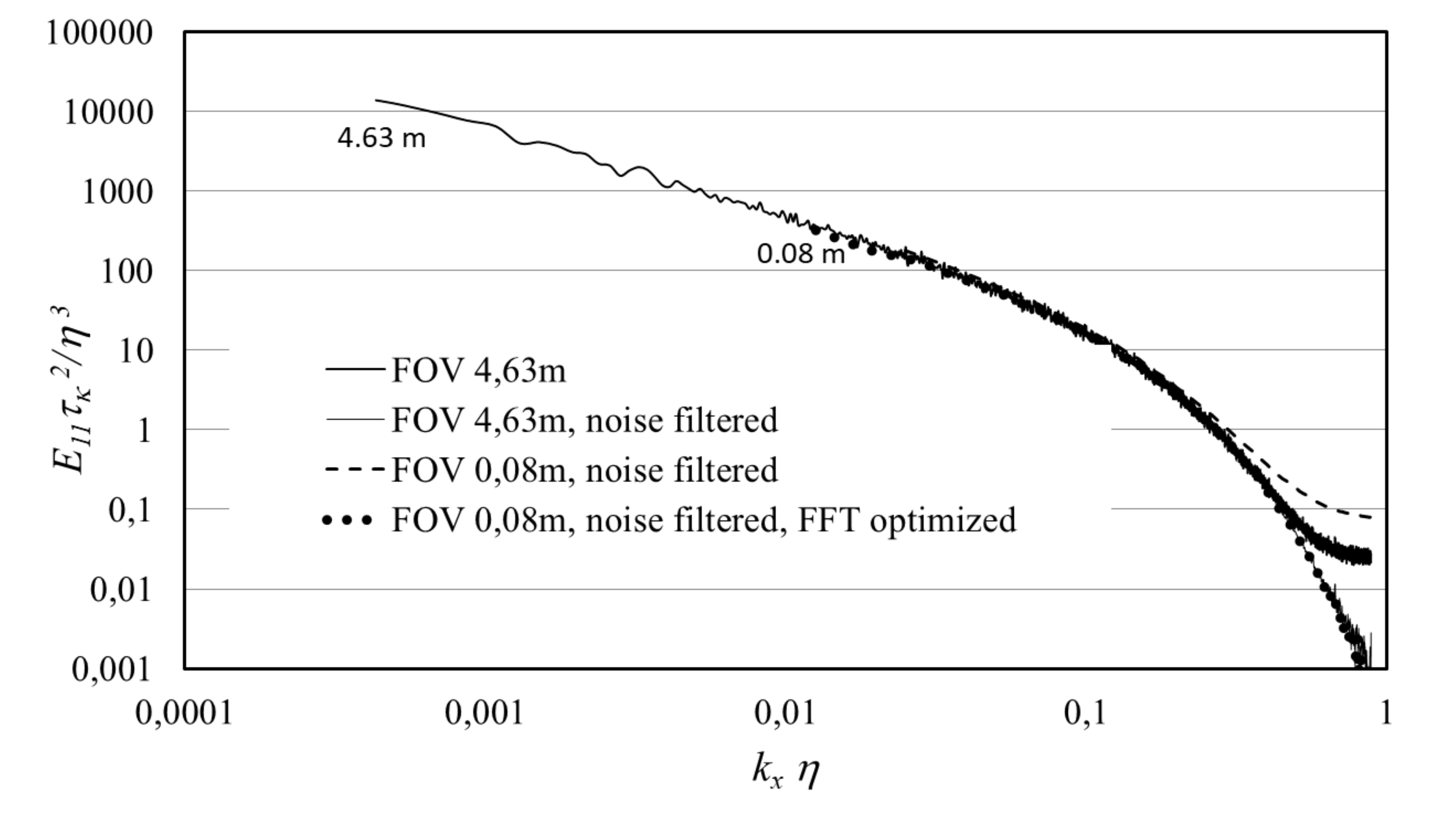}}
	\caption{Simulation of the influence of the limited field of view of PIV on the turbulence spectrum.}
	\label{fig:7}      
\end{figure*}

Figure \ref{fig:7} illustrates this effect on an equivalent HW signal for which the spectrum is first computed on long samples of 2 s (corresponding to a window length of 4.63 m). The presence of the HW electronic white noise is visible at high wave numbers and can be removed by a filter or by substraction.  Also shown on the figure is the spectrum computed on short samples comparable in size to the present PIV field which is 0.12 m in length (based on Taylor hypothesis). The result is quite poor if no special precaution is taken. A significant leakage occurs from the low wave numbers to the high ones. In the present contribution, the data were processed using a methodology equivalent to several Parzen windows (\cite{foucaut04}). Figure \ref{fig:7} shows that this brings a significant improvement to the spectrum for a limited field of view which is close to the original one. The large scales cannot of course be recovered, but the spectrum is reasonable in the range of interest in this experiment.

\begin{figure*}[ht]
	\centering
	\resizebox{0.32\linewidth}{!}{\includegraphics[scale=1]{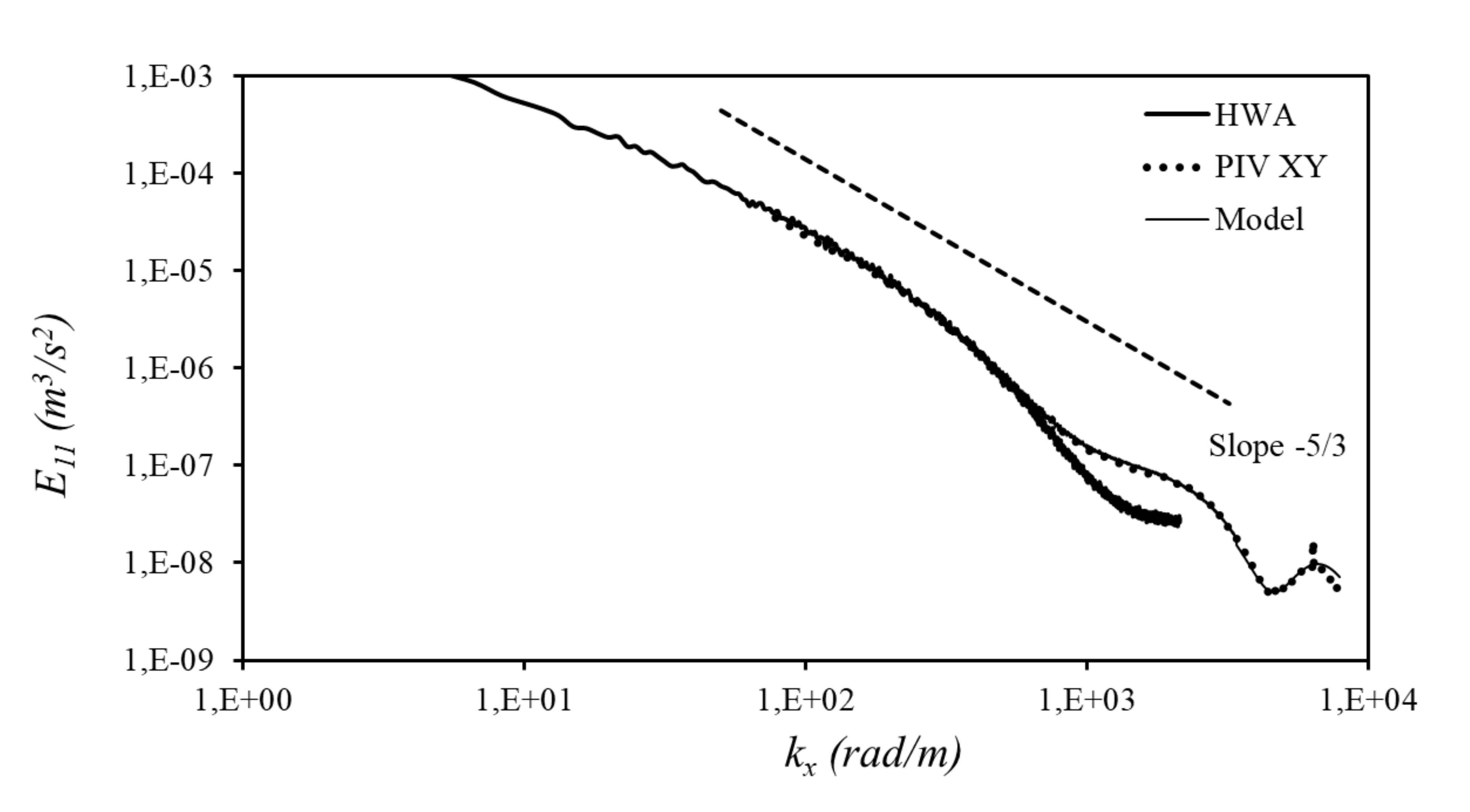}}
	\resizebox{0.32\linewidth}{!}{\includegraphics[scale=1]{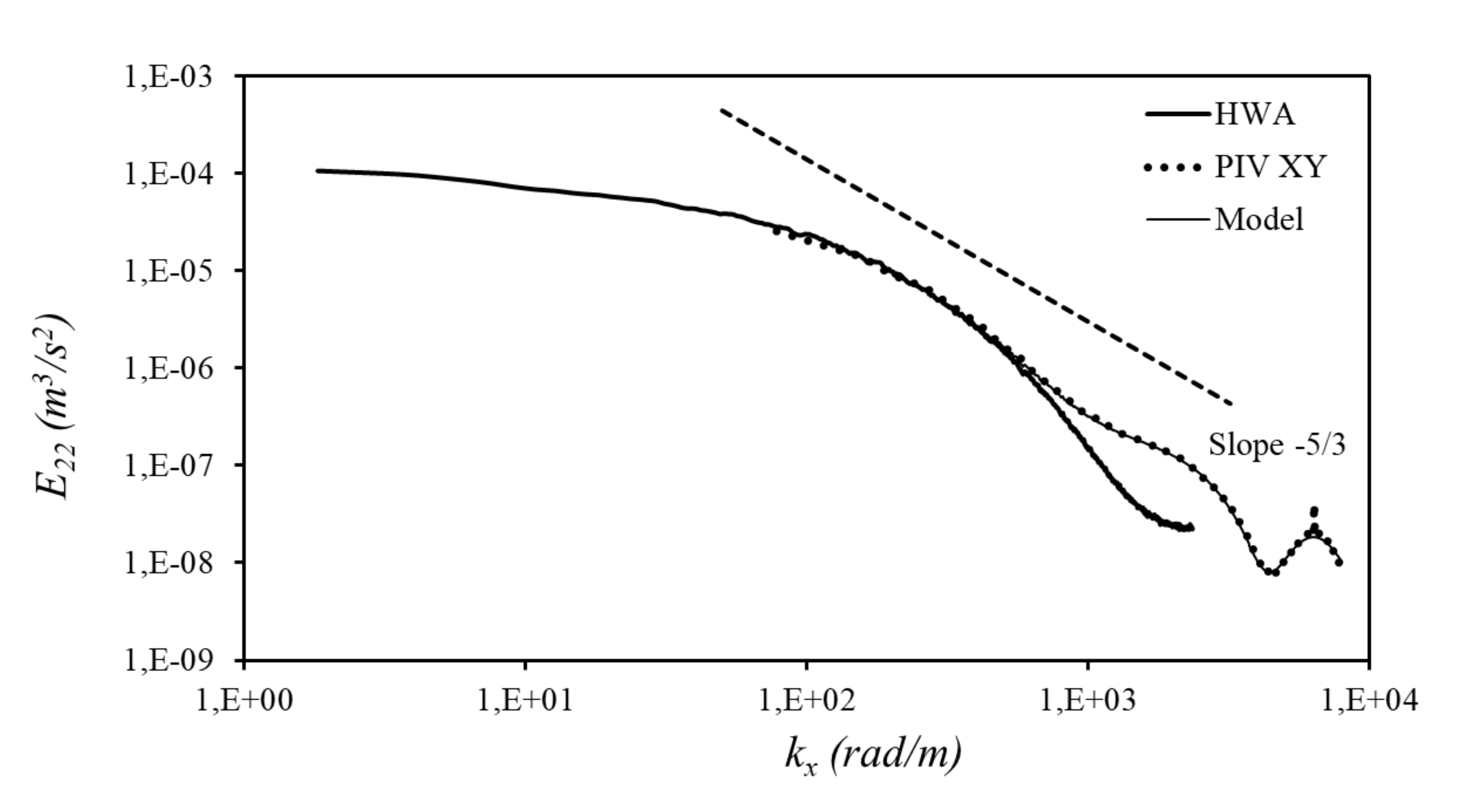}}
	\resizebox{0.32\linewidth}{!}{\includegraphics[scale=1]{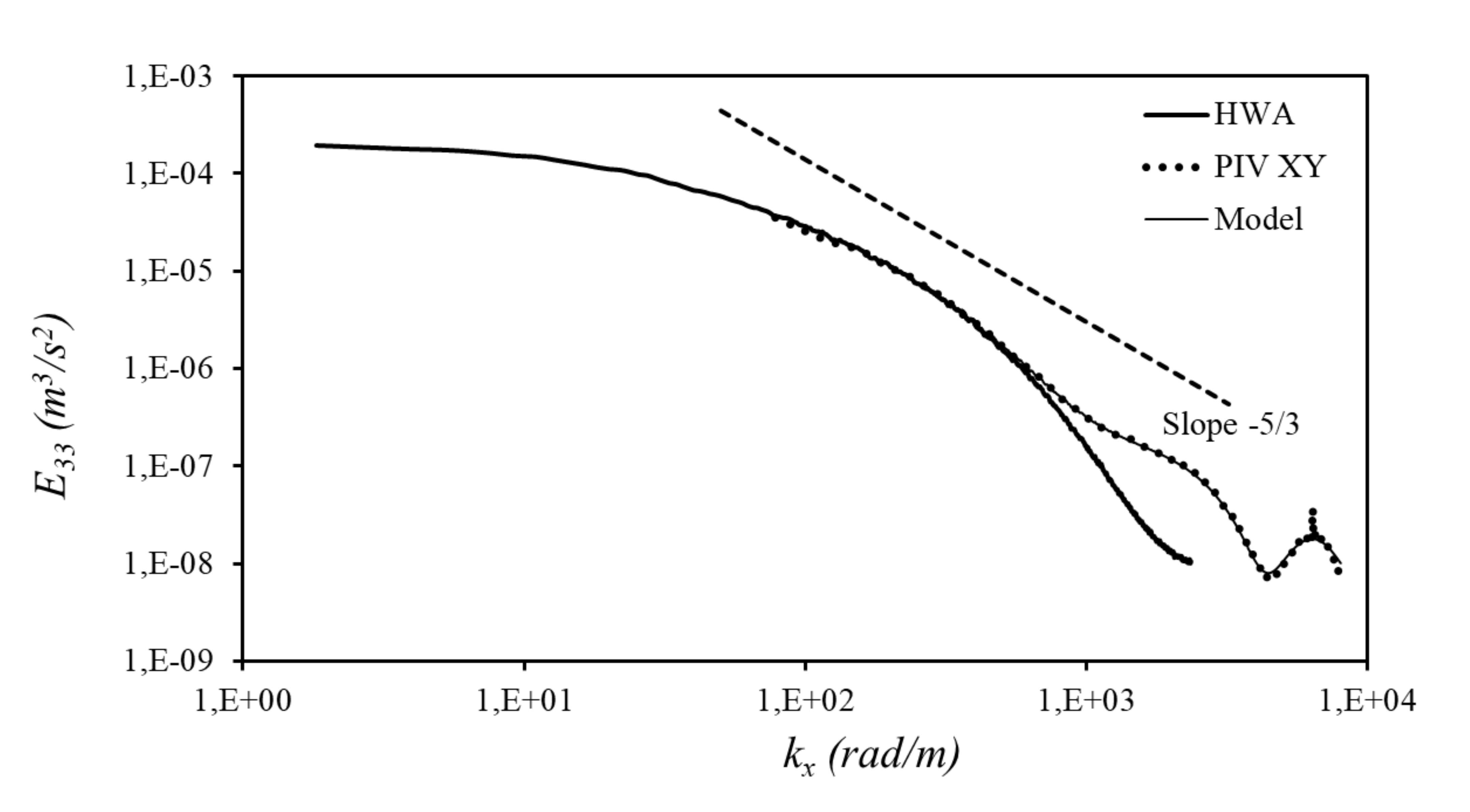}}
	\caption{Comparison of the streamwise and spanwise velocity spectra from PIV and HWA}
	\label{fig:8}      
\end{figure*}

Figure \ref{fig:8} presents now the streamwise, wall-normal and spanwise spectra computed from the streamwise  plane SPIV data at a distance of 200 wall units from the wall. Also shown is the HWA ones using a local Taylor hypothesis for the latter. As can be seen, the agreement is excellent below the cut-off frequence $k_c$ = 900 rad/m. Above this value, the PIV noise and the effect of the IW size are visible. These will be discussed in more detail in the next section. 

The PIV spectrum is compared with the model of~\cite{foucaut04}:

\begin{equation}
  E_{11PIV} = (E_{11hw} + E_{noise})\left(\frac{sin(k W\!S/2)}{k W\!S/2}\right)^2
	\label{eq:03s}
\end{equation}
which takes into account the effect of a constant white noise and of the interrogation window filtering on the original signal.  In this model,  $E_{11hw}$ is the ``real'' turbulence spectrum estimated here by HWA with a Taylor hypothesis, $W\!S$ is the interrogation window size and $E_{noise}$ is obtained by the method described by  \cite{foucaut04}. Note that in \cite{foucaut04} a full PIV approach is proposed if the HWA is not available. 

\begin{figure*}[ht]
	\centering 
	\resizebox{0.85\linewidth}{!}{\includegraphics[scale=1]{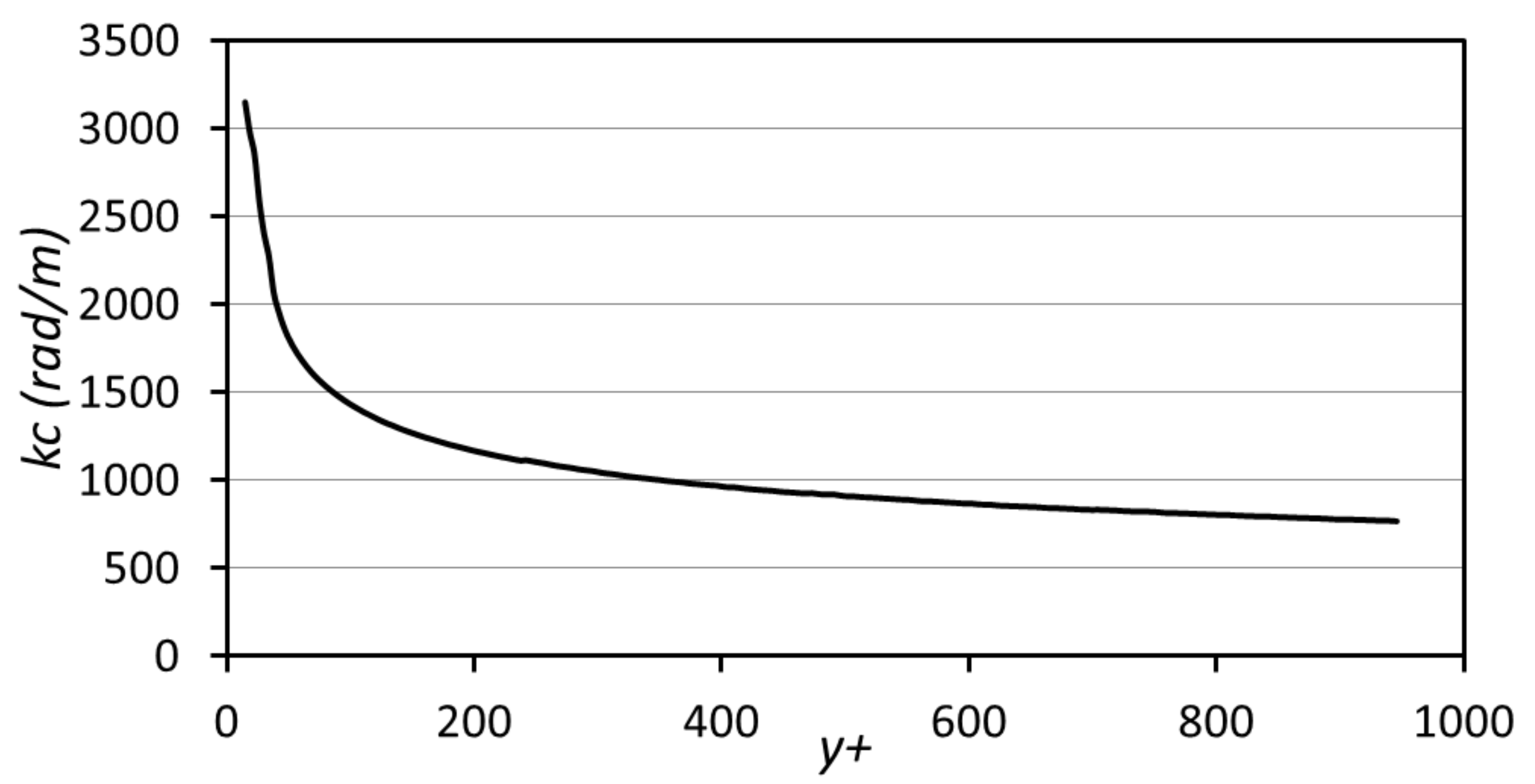}}
	\caption{Cut-off wave number $k_c$ of the PIV measurements as a function of wall distance.}
	\label{fig:9}      
\end{figure*}

By this approach, the PIV noise level $\sigma_1$ on the streamwise component $u_1$ can be quantified. From this noise level, the uncertainty on $u_1$ can be estimated~(\cite{foucaut04}). The same approach can be repeated for the two other components. The corresponding values are given in Table \ref{tab:1} for the streamwise plane. The uncertainties on the components stretched and not  stretched by the stereoscopic angle are of the order of 0.1 pixel and 0.07 pixel, respectively. These values correspond to a PIV measurement of good quality. Of course, the cut-off frequency of the PIV varies with the distance to the wall. This variation is characterized in figure \ref{fig:9}. As will be seen in the next section, it can be taken into account to optimise the derivative filters.

\subsection{Noise assesment}

At the intersection of the two PIV planes, the velocity can be estimated by two independent SPIV systems. 
Along this line, the term $\left\langle u_iu_i \right\rangle$ can be computed in the streamwise or in the spanwise plane or by the product of $u_i$ in each plane. As the two SPIV systems are independent and look at the particles from different angles, the noise can be considered uncorrelated between them. This  combined product  is then quite interesting as it is free from noise. This gives:

\begin{equation}
\langle u_{i}\left|_{(x_1,x_2)}\right. u_{i}\left|_{(x_1,x_2)}\right.\rangle = \langle  u_{i}\left|_{(x_1,x_2)} \right. u_{i}\left|_{(x_3,x_2)} \right.\rangle +  \left(\sigma_{u_{i}} \left|_{(x_1,x_2)} \right.\right)^2
\label{eq:03a}
\end{equation}

\begin{figure*}[ht]
	\resizebox{0.5\linewidth}{!}{\includegraphics[scale=1]{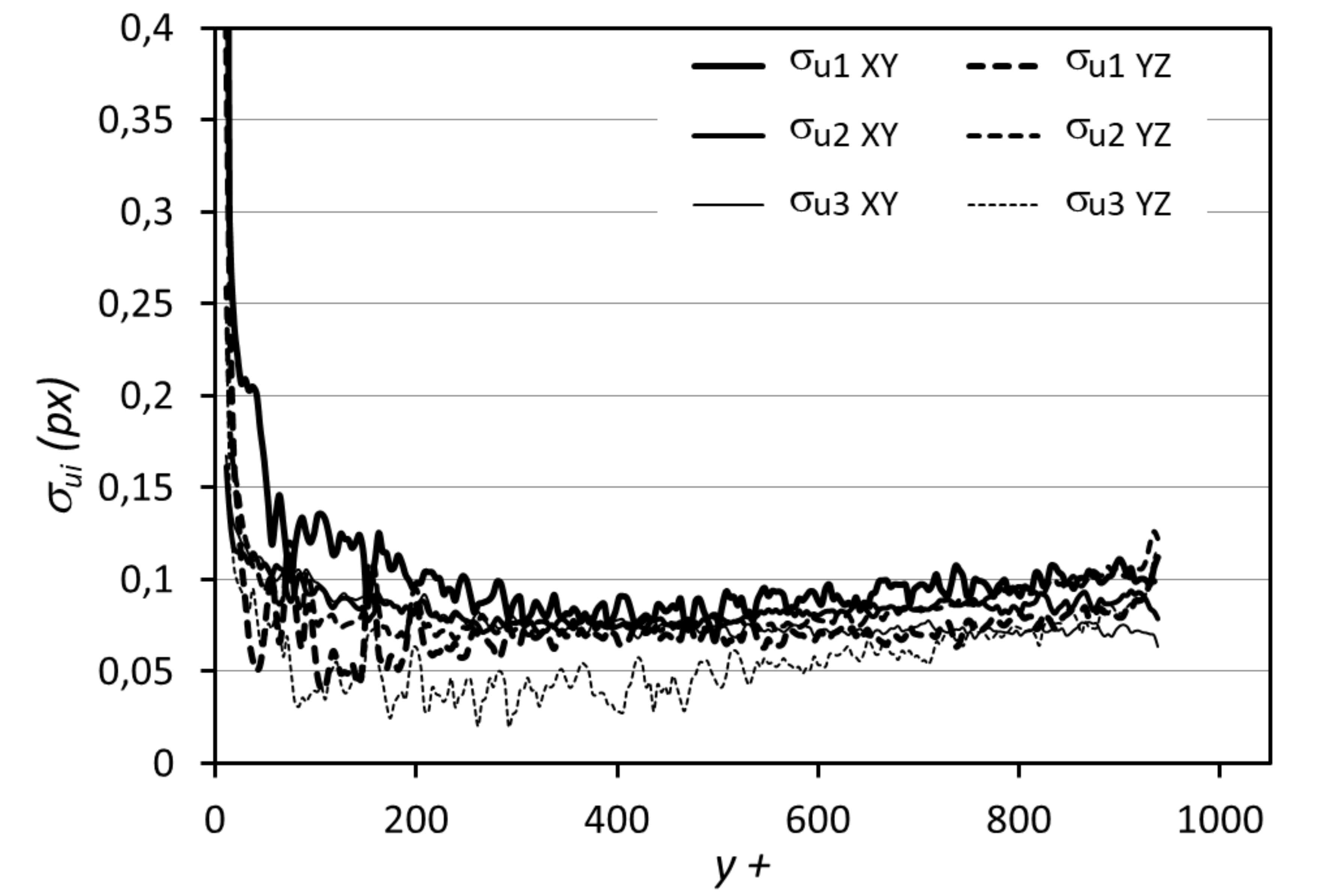}}
	\resizebox{0.5\linewidth}{!}{\includegraphics[scale=1]{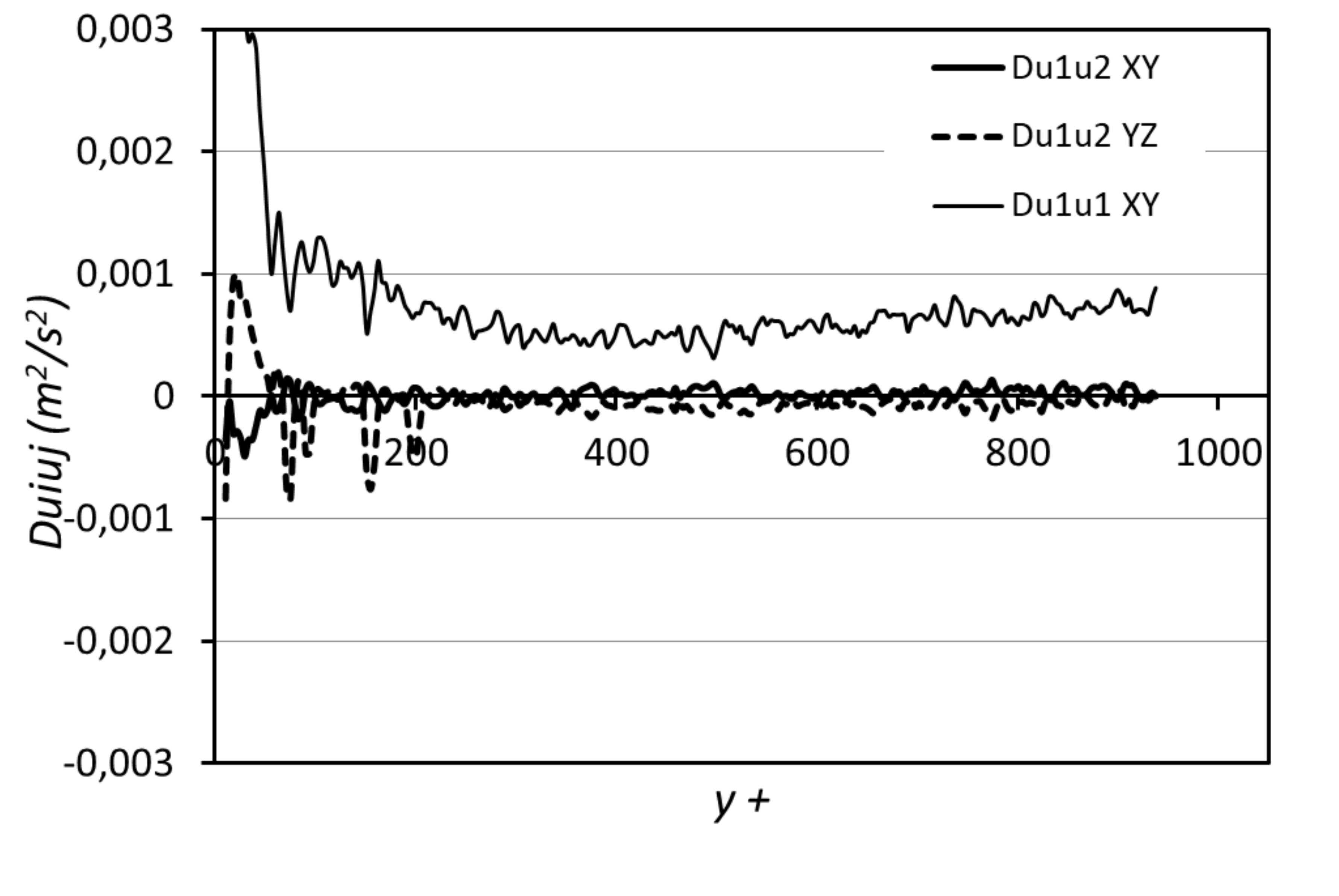}}
	\caption{Estimation of the measurement errors at the intersection line of the two cross-planes.}
	\label{fig:10}      
\end{figure*}

Since the first two terms are measured directly, the uncertainty $\sigma_{u_{i}}$ can be estimated. Figure \ref{fig:10}a shows the profiles of these estimations. In the plane ($x_1,x_2$), the value of $\sigma_{u_i}$ is relatively constant for $y > 5 \;mm$. It increases slowly with increasing with $x_2$ in the plane ($x_3,x_2$), probably due to an increasing  loss of pairs in such a cross-flow plane SPIV when $x_2$ (and the mean velocity) increases  (v.\ ~\cite{foucaut14}). Close to the wall, the uncertainties increases because of the strong mean velocity gradient even if a deformation technique is used. 

An average of the values for $x_2 > 5 \;mm$ was computed for each component and each plane. These values are included in Table \ref{tab:1}. For the plane ($x_1,x_2$) they are of the same order as the ones deduced from the spectral analysis. For the plane ($x_3,x_2$) the values are smaller showing a very good quality of measurement for a plane normal to the flow. The errors on the $u_3$ component  are a little smaller (0.036 pix) than the other two because this component is not stretched in this plane. The measurement uncertainties are finally smaller than 0.1 pixel everywhere, which is representative of a good quality in both planes. 

By analogy with equation (\ref{eq:03a}), we can define $\Delta u_iu_j$ by:
\begin{equation}
\Delta u_iu_j = \left\langle \left.u_{i}\right|_{(x_1,x_2)}\left. u_{j}\right|_{(x_1,x_2)} \right\rangle - \left\langle \left. u_{i}\right|_{(x_1,x_2)} \left. u_{j}\right|_{(x_3,x_2)} \right\rangle 
\label{eq:03b}
\end{equation}
Figure \ref{fig:10}b show a comparison of $\Delta u_1u_2$ for each plane and $\Delta u_1u_1 = \left(  \left. \sigma_{u_{i}} \right|_{(x_1,x_2)} \right)^2$).
The values of $\Delta u_1u_2$ are close to zero for each plane. This is due to the fact that the noise from two different components is not correlated which is not the case for $\Delta u_1u_1$.

\begin{table*}[h]
\caption{Noise level and uncertainties of the SPIV measurement, $E_{noise}$ is the noise level (see \cite{foucaut04}), $W\!S$ is the PIV window size and $\sigma_u$ is the PIV measurement uncertainties}
\label{tab:1}       
\begin{center}
\begin{tabular}{lllll}
\hline\noalign{\smallskip}
Plane & Component & $E_{noise} (m^3/s^2)$ & $\sigma_u$ SP (pix) & $\sigma_u$ Dif. (pix)   \\
\noalign{\smallskip}\hline\noalign{\smallskip}
$x_1,x_2$ & $u_1$ & $1.0\;10^{-7}$  & 0.076 & 0.092\\
$x_1,x_2$ & $u_2$ & $2.2\;10^{-7}$  & 0.069 & 0.078\\
$x_1,x_2$ & $u_3$ & $2.1\;10^{-7}$  & 0.068 & 0.079\\
$x_3,x_2$ & $u_1$ &  &  & 0.064\\
$x_3,x_2$ & $u_2$ &  &  & 0.073\\
$x_3,x_2$ & $u_3$ &  &  & 0.036\\
\noalign{\smallskip}\hline
\end{tabular}
\end{center}
\end{table*}

\section{Derivatives computation}
\label{sec:3}

The next step toward the characterization of the TKE dissipation is to evaluate the spatial derivatives of the velocity field.  Before entering into the details of the approach used here, it is of interest to try to understand and characterize the effect the PIV interrogation window (IW) on the resolution of the dissipative scales.

\subsection{Interrogation window size and resolution}

 For the purpose of understanding the effect of the IW size on the spatial resolution, it is possible to use the HW signal which was recorded on long samples (2 s) and with a good temporal resolution (2.2 kHz). On the basis of a Taylor hypothesis this resolution corresponds at $y^+ = 200$ to 0.88 mm that is 2.5 Kolmogorov units and at $y^+ = 45$ to 0.72 mm that is 2.2 Kolmogorov units. With such a signal, it is possible to simulate the effect of the IW. Before doing that, it is necessary to remove from the HW spectrum the low white noise so that the $k^2E(k)$ dissipation spectrum goes effectively to zero at high wave numbers. 
 
 \begin{figure*}[ht]
 	\centering 
 	\resizebox{0.45\linewidth}{!}{\includegraphics[scale=1]{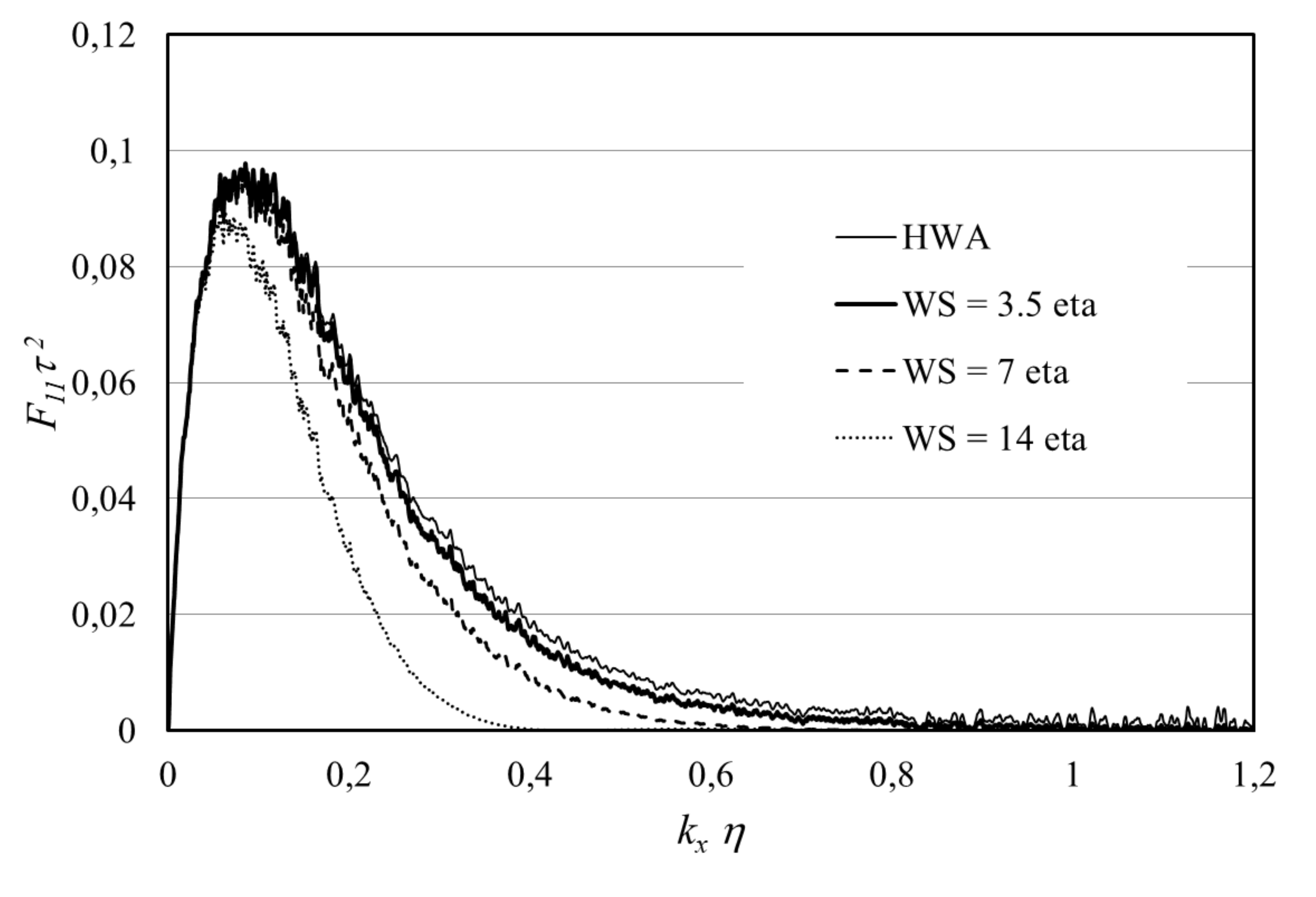}}
 	\resizebox{0.45\linewidth}{!}{\includegraphics[scale=1]{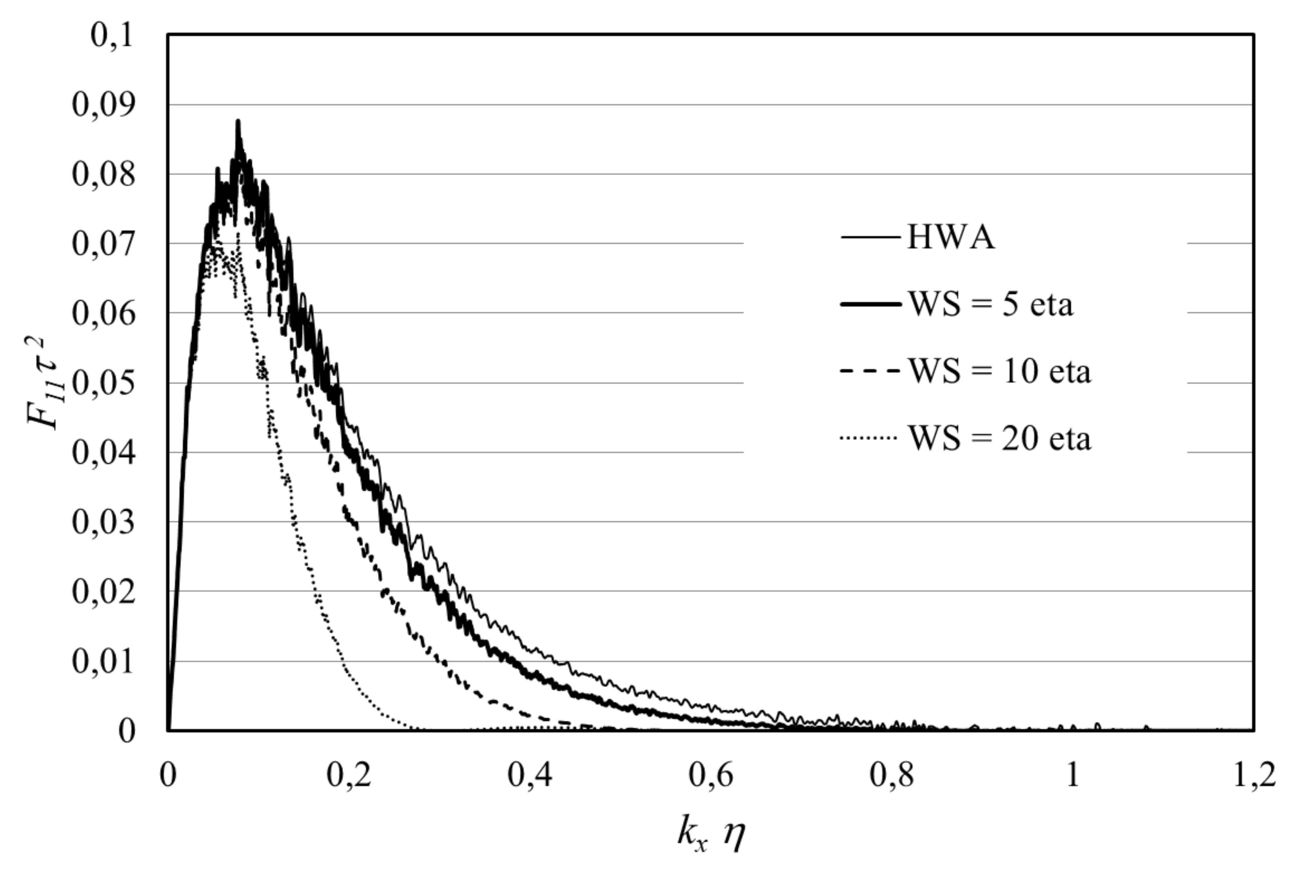}}\\
 	\hspace{3 cm}(a)   \hspace{5 cm}   (b)
 	\caption{Simulation of the effect of the PIV interrogation window size IW on the attenuation of the dissipation spectra at two wall distances $y^+ = 45$ \& $200$.}
 	\label{fig:11}      
 \end{figure*}

 Figure \ref{fig:11} provides the dissipation spectrum $F_{11}$ computed from this HW signal at two wall distances $y^+ = 45$ \& $200$ with 3 different values of the interrogation window given in Kolmogorov units (the HW probe was 0.5 mm in length, that is about 1.5 Kolmogorov units, and 2.5 microns in diameter). For comparison, at $y^+ = 45$ the PIV IW is about 5 Kolmogorov units and at $y^+ = 200$ it is about 3. As expected, the reduction of the size of the interrogation window reduces the spatial filtering and provides a better and better estimation of the spectrum. The curves corresponding to  5 and 3 Kolmogorov units at $y^+ = 45$ and $200$ respectively are the  closest to the 1.4 mm PIV IW. These are representative of the best possible estimation that the PIV could  provide if it was free of noise. With this size of window, the dissipation obtained by integrating the spectrum is about 10 \%  lower than that provided by the full HW spectrum. These results will be used later to evaluate the derivative estimation approach used in the present study. Note that these estimates are close to those of \cite{lavoie07} computed for PIV using DNS spectra and a three-dimensional filter. 
 
\subsection{Derivatives computation approach}

In the present contribution, the derivatives were computed using the second-order centered-difference scheme proposed by \cite{foucaut02}. The basic scheme is :
\begin{equation}
	\frac{\partial u_i}{\partial x_j} = \frac{u_i^{p+n}-u_i^{p-n}}{2n\Delta x_j}+\frac{n^2 \Delta x_j^2}{2} \frac{\partial^2 u_i}{\partial x_j^2}+\sigma_\frac{\partial u_i}{\partial x_j}
	\label{eq:03}
\end{equation}
where $\Delta x_j$ is the grid spacing along $x_j$. The first term on the right-hand side corresponds to a second order derivative scheme at grid point $p$, the second term  corresponds to the truncation error. Increasing the value of n decreases the cut-off frequency of the scheme. The term $\sigma_\frac{\partial u_i}{\partial x_j} = \alpha \frac{\sigma_{u_i}}{n \Delta x_j}$ is the error due to the noise on the derivative where $\sigma_{u_i}$ is the random measurement error on the velocity and $\alpha$ is the quadratic sum  of the coefficients used to compute the derivative. Following \cite{foucaut02}, in the case of a second-order centered-difference scheme,  $\alpha = 1/\sqrt{2}$. As can be seen, $\sigma_\frac{\partial u_i}{\partial x_j}$  is inversely proportional to the spacing $n \Delta x_j$  between the two points used to compute the derivative (\cite{foucaut02}). This term is representative of the noise amplification. The higher n, the lower the noise. The choice of n is consequently a compromise between maximizing the cut-off frequency and minimizing the noise. Table \ref{tab:2} gives the values of the cut-off frequency and noise level for $1 \leq n \leq 4$. For n = 3 the cut-off frequency of the derivative filter is 925 rad/s which is of the order of the PIV cut-off frequency at $y+ = 200$ as shown in Figure \ref{fig:8}. 
%

\begin{table*}[h]
\caption{Characteristics of the derivative schemes}
\label{tab:2}       
\begin{center}
\begin{tabular}{lll}
\hline\noalign{\smallskip}
n & $k_c$ (rad/m) & $\sigma_\frac{\partial u_i}{\partial x_j}$   \\
\noalign{\smallskip}\hline\noalign{\smallskip}
1 & 2800  & 45.44\\
2 & 1350 & 22.72\\
3 & 925  & 15.15\\
4 & 700  & 11.36\\
\noalign{\smallskip}\hline
\end{tabular}
\end{center}
\end{table*}

\begin{figure*}[ht]
	\resizebox{0.85\linewidth}{!}{\includegraphics[scale=1]{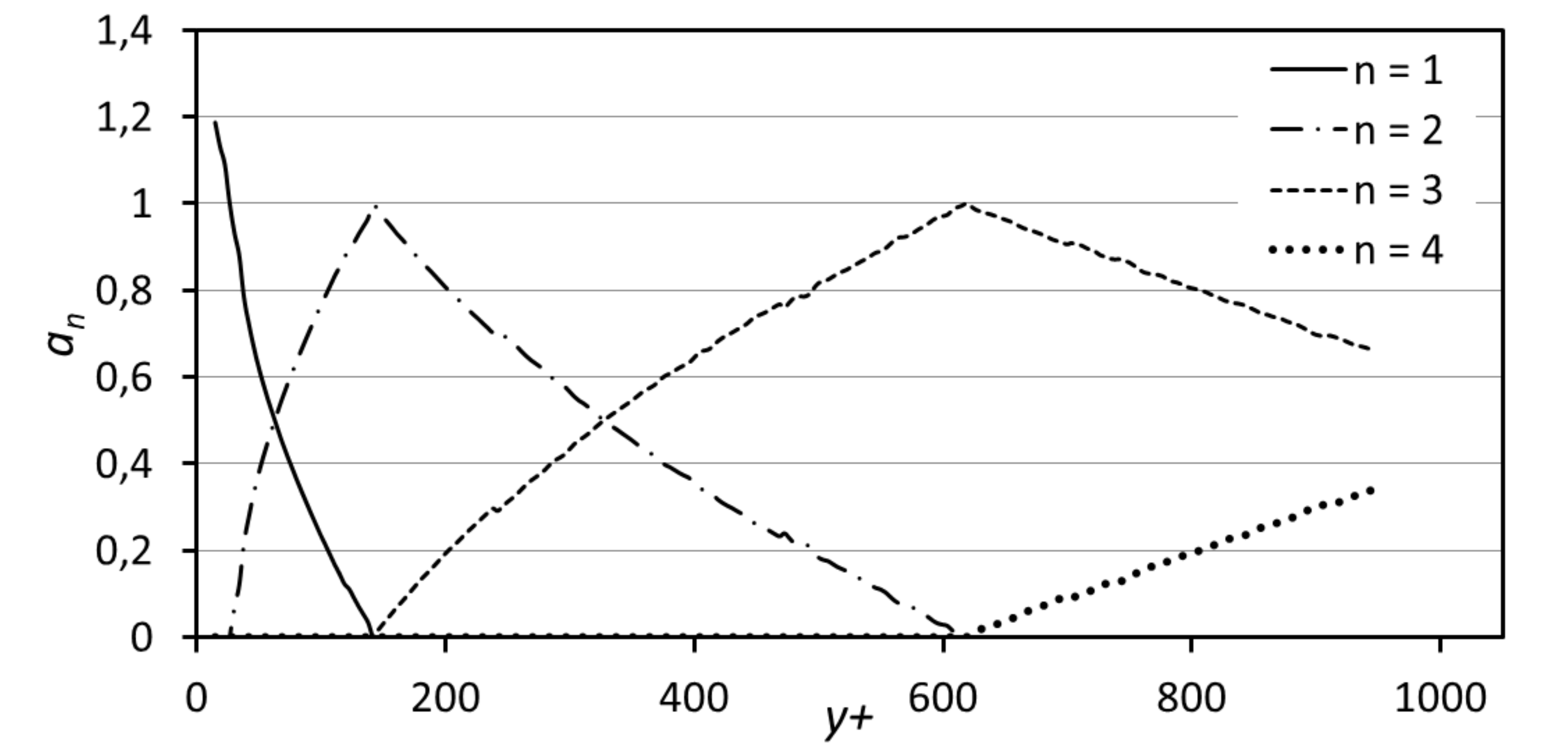}}
	\caption{Evolution of the coefficients $a_n$ of equation  (\ref{eq:04}) versus the wall distance.}
	\label{fig:12}      
\end{figure*}

\begin{figure*}[ht]
	\resizebox{0.85\linewidth}{!}{\includegraphics[scale=1]{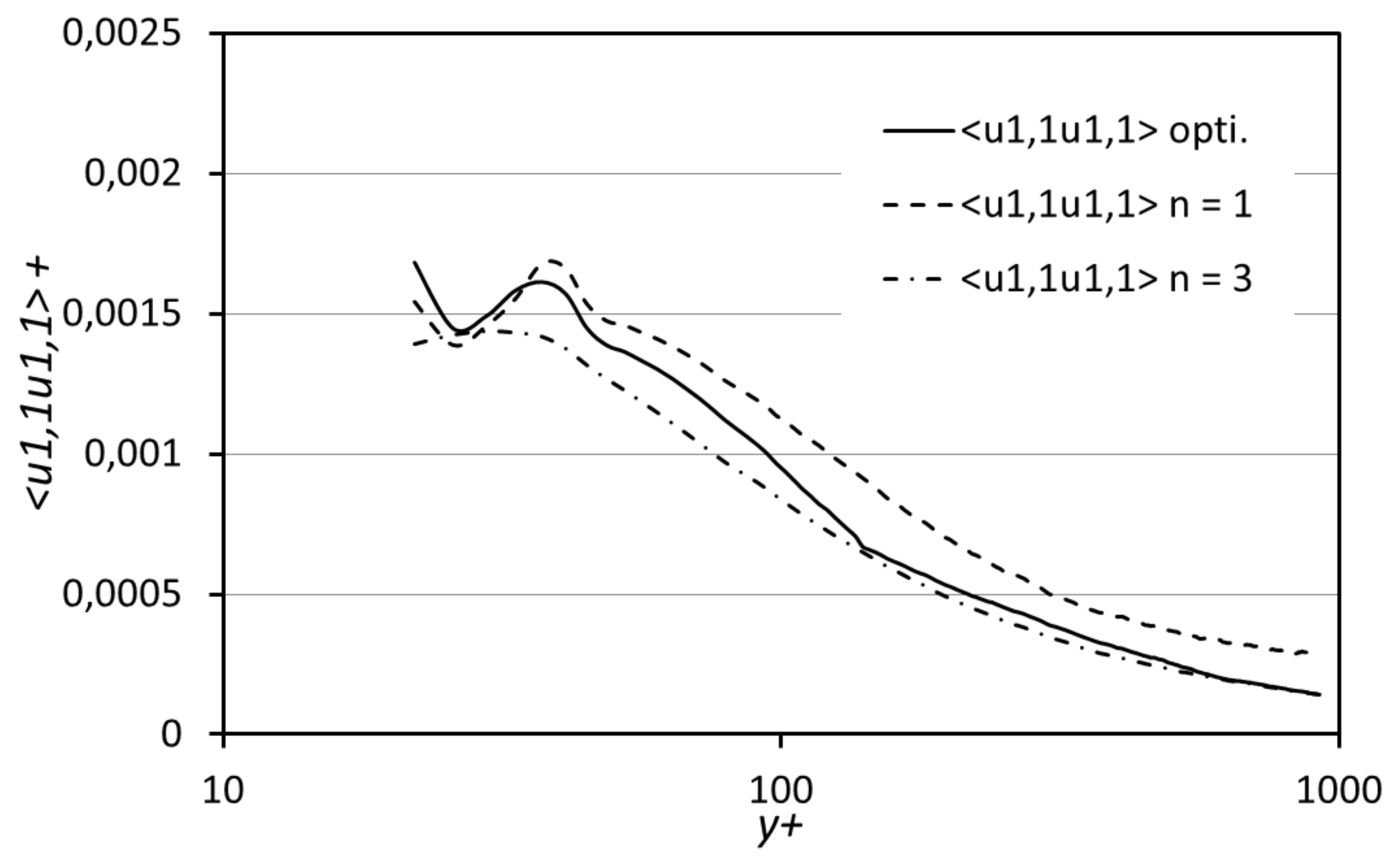}}
	\caption{Profiles of $\left\langle \frac{\partial u_1}{\partial x_1}\frac{\partial u_1}{\partial x_1} \right\rangle$ computed with n = 1, n = 3 and n optimised.}
	\label{fig:13}      
\end{figure*}

In practice, the measurement noise on the velocities is relatively constant, but the signal level decreases when the wall distance increases. Consequently, to keep a relatively constant signal-to-noise ratio, the cut-off frequency of the derivative scheme has to be reduced. To accomplish that, the derivative can be computed from a combination of central-difference schemes at different orders:
\begin{equation}
	\frac{\partial u_i}{\partial x_j} = \sum_{n=1,4}{a_n\frac{u_i^{p+n}-u_i^{p-n}}{2n\Delta x_j}},
	\label{eq:04}
\end{equation}
where the coefficients $a_n$ are optimised as a function of the wall distance to obtain the best cut-off frequencies. The resulting values of $a_n$ in the present case are plotted in Figure \ref{fig:12}. Figure \ref{fig:13} compares the profiles of $\left\langle \frac{\partial u_1}{\partial x_1}\frac{\partial u_1}{\partial x_1} \right\rangle$ computed with n = 1, n = 3 and n optimised.  As can be seen, the derivative with the optimisation is in agreement with n = 1 close to the wall and with n = 3 further away. 

\begin{figure*}[ht]
	\resizebox{0.5\linewidth}{!}{\includegraphics[scale=1]{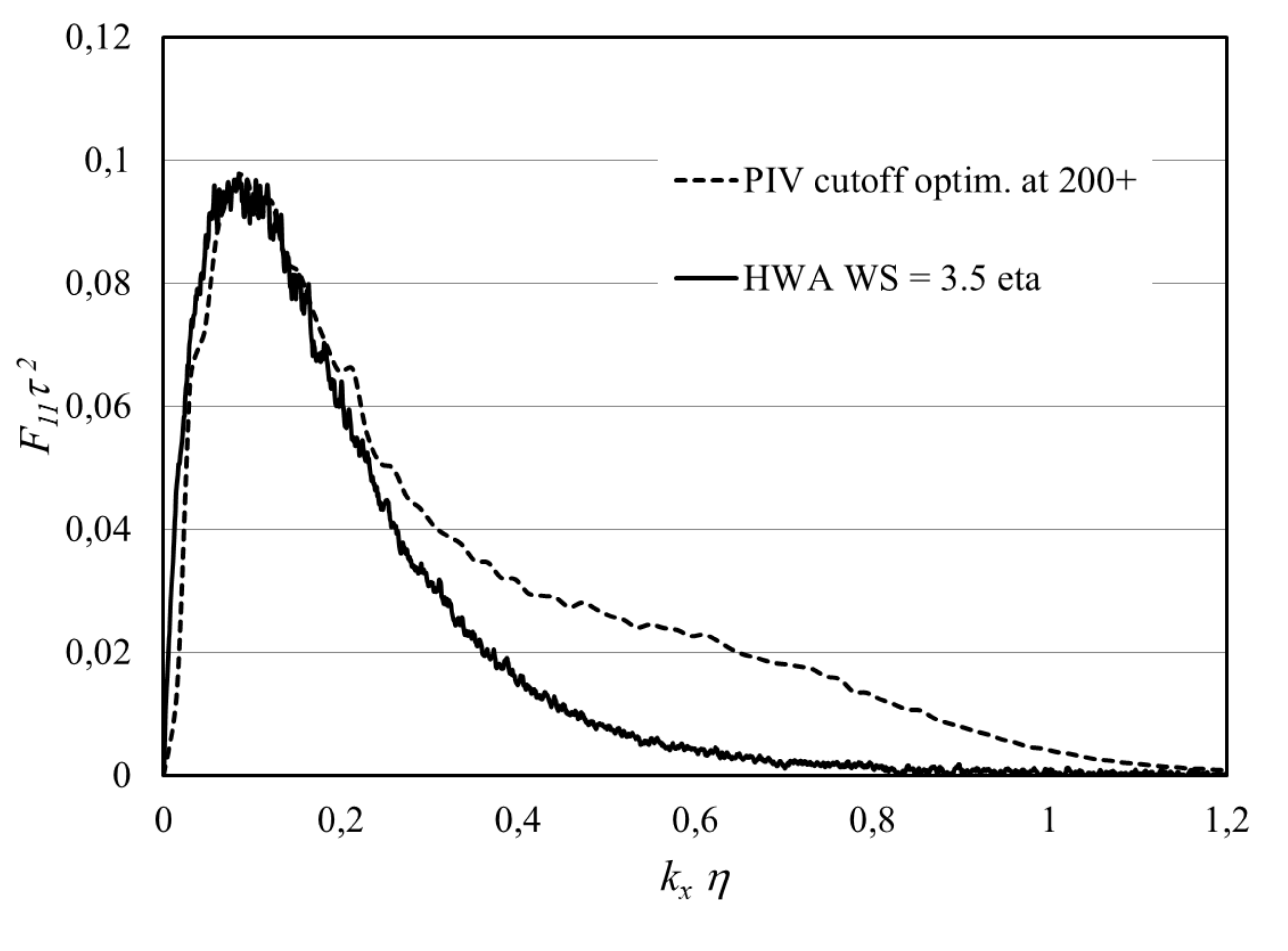}}
	\resizebox{0.5\linewidth}{!}{\includegraphics[scale=1]{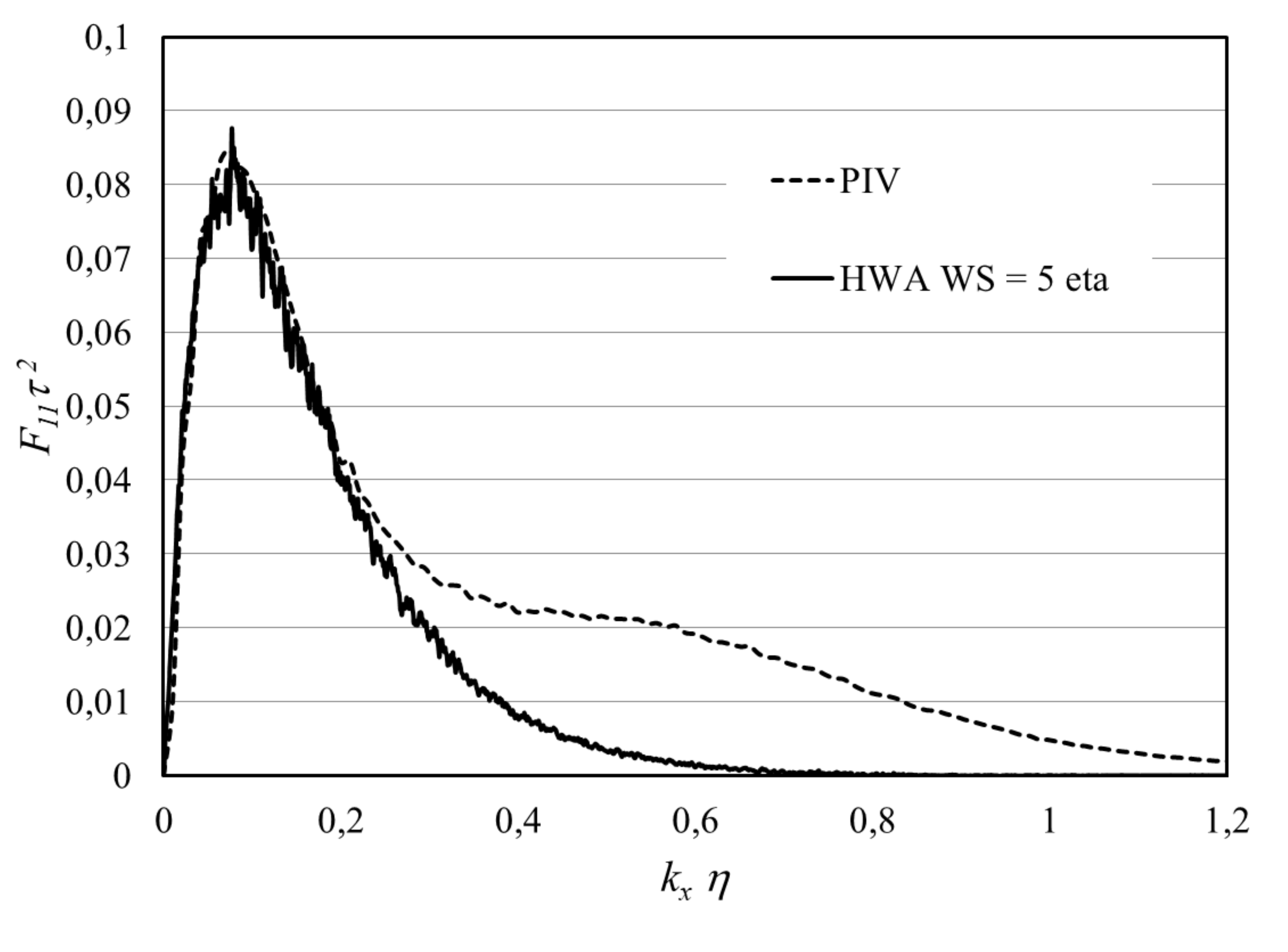}}
	\caption{Spectrum of the term $\left\langle \frac{\partial u_1}{\partial x_1}\frac{\partial u_1}{\partial x_1} \right\rangle$ compared with a HW simulation at $y^+ = 45$ \& $200$.}
	\label{fig:14}      
\end{figure*}

Figure \ref{fig:14} shows a comparison  at $y^+ = 45$ and $200$ of the spectrum of $\frac{\partial u_1}{\partial x_1}$ obtained by the above procedure at $Re_\tau = 2300$ with the $k^2E(k)$ HW spectrum shown already in figure \ref{fig:11}. The latter has had the white noise removed and has been computed  on windows of 0.12 m and simulating an IW of 3.5 and 5 Kolmogorov respectively. As can be observed, the compromise chosen for the IW and the derivative stencil represents well the peak of the spectrum, but introduces some noise at high frequency. Of course, it would be possible to optimise the derivative filter to bring the PIV spectrum closer to the HW spectrum at high frequency. But this would be at the expense of some signal at the lower frequencies, as the filter does not really distinguish between signal and noise. The choice here was to make a compromise giving a signal-to-noise ratio  of one at a cut-off frequency which is close to the velocity fluctuations cut-off frequency. As will be seen below, this remaining noise can be assessed and removed from the derivative moments, thanks to the cross-planes configuration and the use of the continuity equation.

\section{Derivative moments}
\label{sec:4}

In order to estimate the dissipation it is required to obtain the derivative moments which appear in equation (\ref{eq:02}). 

\subsection{Three methodogies}

Thanks to the above, there are three independent methods for computing the derivatie moments. As for the Reynolds stresses, each moment of the type $\left\langle \frac{\partial u_i}{\partial x_2}\frac{\partial u_i}{\partial x_2} \right\rangle$ can be computed in the streamwise or in the spanwise plane or by the product of derivative in each plane. The last possibility gives again a result which is free from noise as the noise from the two systems are uncorrelated. By a simple difference, the noise can be estimated:

\begin{equation}
\sigma_{\left.\frac{\partial u_{i}}{\partial x_2}\right|_{(x_1,x_2)}}^2 = \left\langle \left.\frac{\partial u_{i}}{\partial x_2}\right|_{(x_1,x_2)}\left.\frac{\partial u_{i}}{\partial x_2}\right|_{(x_1,x_2)} \right\rangle - \left\langle \left.\frac{\partial u_{i}}{\partial x_2}\right|_{(x_1,x_2)} \left.\frac{\partial u_{i}}{\partial x_2}\right|_{(x_3,x_2)} \right\rangle 
\label{eq:06}
\end{equation}

A second possibility to estimate the noise is to multiply each normal derivative by the continuity equation:

\begin{eqnarray}
\left\langle \frac{\partial u_1}{\partial x_1}\frac{\partial u_1}{\partial x_1} \right\rangle + \left\langle \frac{\partial u_1}{\partial x_1}\frac{\partial u_2}{\partial x_2} \right\rangle + \left\langle \frac{\partial u_1}{\partial x_1}\frac{\partial u_3}{\partial x_3} \right\rangle = \sigma_{\left.\frac{\partial u_{1}}{\partial x_1}\right|_{(x_1,x_2)}}^2 \nonumber \\
\left\langle \frac{\partial u_2}{\partial x_2}\frac{\partial u_1}{\partial x_1} \right\rangle + \left\langle \frac{\partial u_2}{\partial x_2}\frac{\partial u_2}{\partial x_2} \right\rangle + \left\langle \frac{\partial u_2}{\partial x_2}\frac{\partial u_3}{\partial x_3} \right\rangle = \sigma_{\left.\frac{\partial u_{2}}{\partial x_2}\right|_{(x_1,x_2)}}^2 \nonumber \\
\left\langle \frac{\partial u_3}{\partial x_3}\frac{\partial u_1}{\partial x_1} \right\rangle + \left\langle \frac{\partial u_3}{\partial x_3}\frac{\partial u_2}{\partial x_2} \right\rangle + \left\langle \frac{\partial u_3}{\partial x_3}\frac{\partial u_3}{\partial x_3} \right\rangle = \sigma_{\left.\frac{\partial u_{3}}{\partial x_3}\right|_{(x_3,x_2)}}^2
\label{eq:06c}
\end{eqnarray}
Note only the diagonal terms are squared, and that the other terms are relatively noise-free  So the latter can be used to calculate a noise-free estimate of the squared terms. As can be seen, the noise $\sigma_{\frac{\partial u_{i}}{\partial x_i}}^2$ can be estimated from each system.

A third possibility is to follow \cite{foucaut02} and to estimate the noise by:
\begin{equation}
\sigma_{\left.\frac{\partial u_{1}}{\partial x_1}\right|_{(x_1,x_2)}}^2 = \alpha^2 \frac{\sigma_{u_1}^2}{\Delta_x^2} \sum_{n=1,4}{ \frac{a_n^2}{n^2}} 
\label{eq:06n}
\end{equation}

\subsection{Comparison of the various methods}

\begin{figure*}[ht]
	\resizebox{0.85\linewidth}{!}{\includegraphics[scale=1]{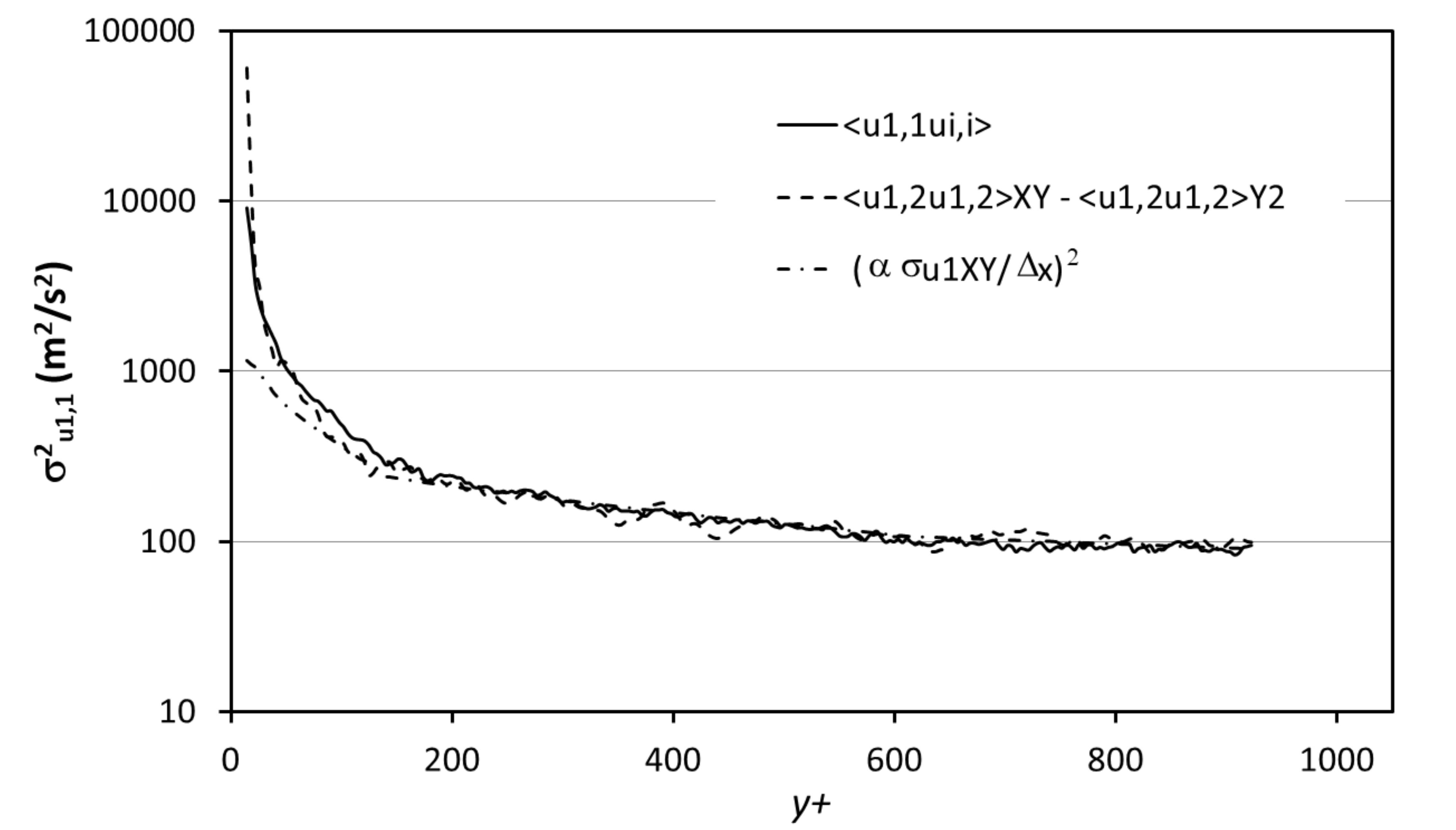}}
	\caption{Estimation of the random noise error on the derivative $\left\langle \frac{\partial u_1}{\partial x_1}\frac{\partial u_1}{\partial x_1} \right\rangle$ by the 2 plane denoising approach and by the continuity.}
	\label{fig:15}      
\end{figure*}

Figure \ref{fig:15} shows the noise error on the derivative $\left\langle \frac{\partial u_1}{\partial x_1}\frac{\partial u_1}{\partial x_1} \right\rangle$ obtained from the cross-planes denoising approach (equation (\ref{eq:06})), the continuity (equation (\ref{eq:06c})) and the model of equation (\ref{eq:06n}). These three estimations give sensibly the same results.  Therefore they can be used to quantify and remove the noise from the derivative moments. For example, the signal-to-noise ratio for $\left\langle \frac{\partial u_1}{\partial x_1}\frac{\partial u_1}{\partial x_1} \right\rangle$ is

\begin{equation}
\nonumber
\sigma_{\left.\frac{\partial u_{1}}{\partial x_1}\right|_{XY}} / \sqrt{\left\langle \frac{\partial u_1}{\partial x_1}\frac{\partial u_1}{\partial x_1} \right\rangle} \approx 0.6
\label{snr}
\end{equation}
close to the wall and $0.9$ far from the wall for the present configuration. Note that it  was about 0.4 and 0.6 at $y^+ = 45$ and $200$ respectively in figure \ref{fig:14} before removing the noise.  

\begin{figure*}[ht]
	\resizebox{0.85\linewidth}{!}{\includegraphics[scale=1]{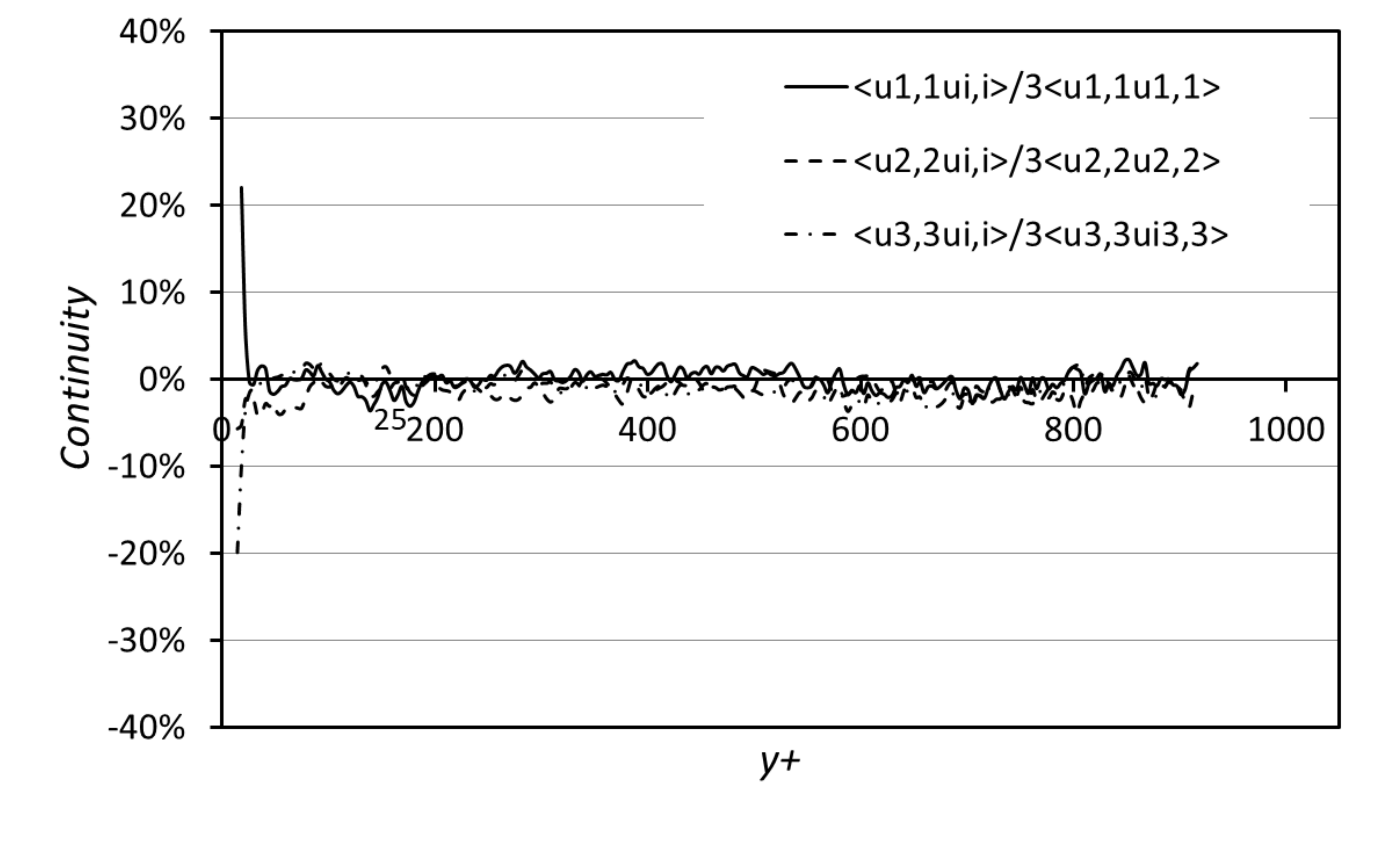}}
	\caption{Profiles of $ {\left\langle \frac{\partial u_k}{\partial x_k}\frac{\partial u_i}{\partial x_i} \right\rangle}/{\left\langle \frac{\partial u_k}{\partial x_k}\frac{\partial u_k}{\partial x_k} \right\rangle}$ with k = 1, 2 and 3 to verify the continuity.}
	\label{fig:16}      
\end{figure*}

An estimation of the random error can also be obtained after this noise removal by using the continuity equation. Figure \ref{fig:16} shows the profiles of $ {\left\langle \frac{\partial u_k}{\partial x_k}\frac{\partial u_i}{\partial x_i} \right\rangle}/{\left\langle \frac{\partial u_k}{\partial x_k}\frac{\partial u_k}{\partial x_k} \right\rangle}$ for k = 1, 2 and 3 which should verify the continuity. The fluctuations are on the order of 10\%. As these data corresponds to the sum of three terms which are random, their value can be divided by $\sqrt{3}$, which gives a random error of the order of 6\% on each term. Note that our above simulation with the hot-wire signal gave about 10\%. 

\subsection{The derivative moment data}
\label{sec:5}

\begin{figure*}[ht]
	\resizebox{0.76\linewidth}{!}{\includegraphics[scale=1]{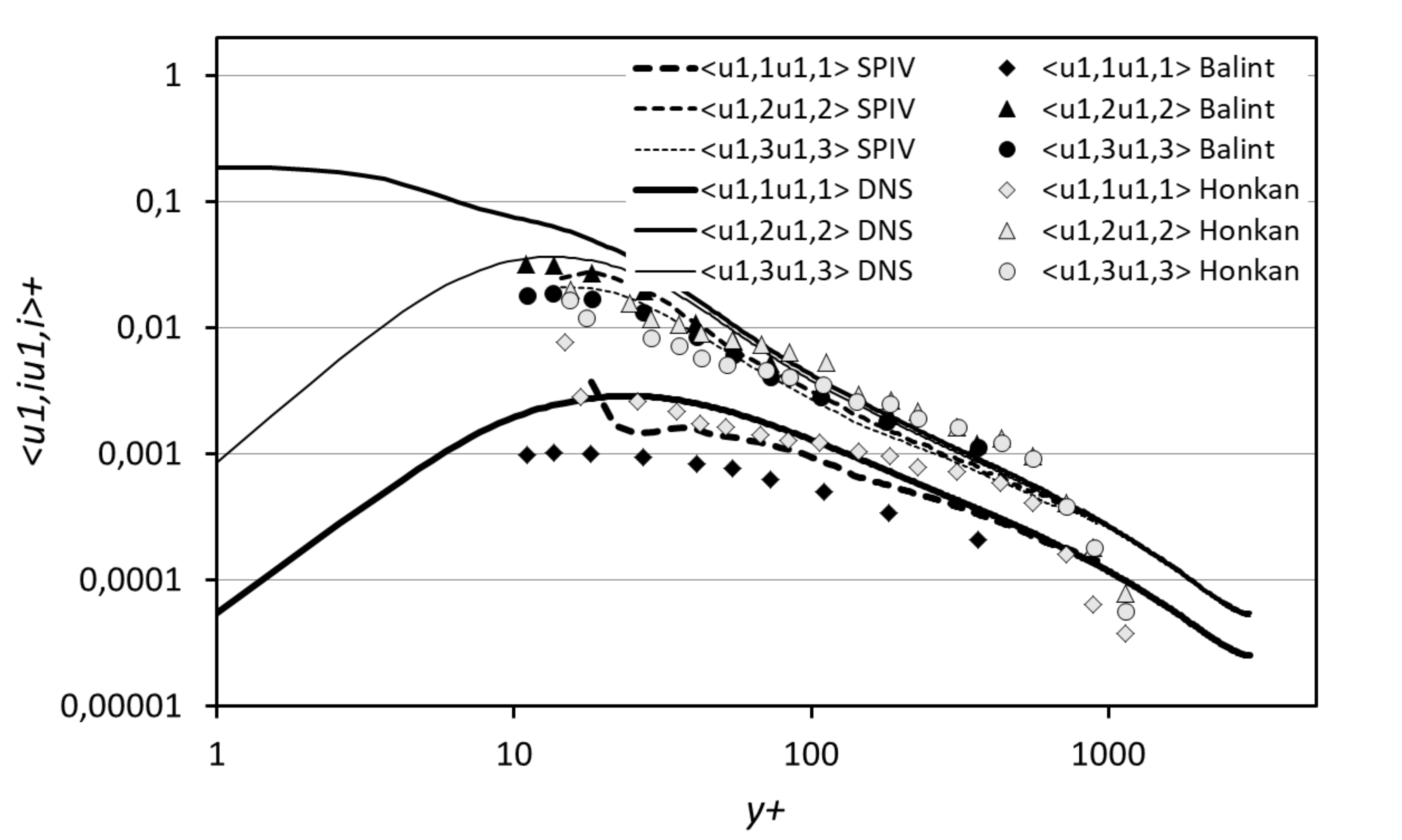}}
	\resizebox{0.76\linewidth}{!}{\includegraphics[scale=1]{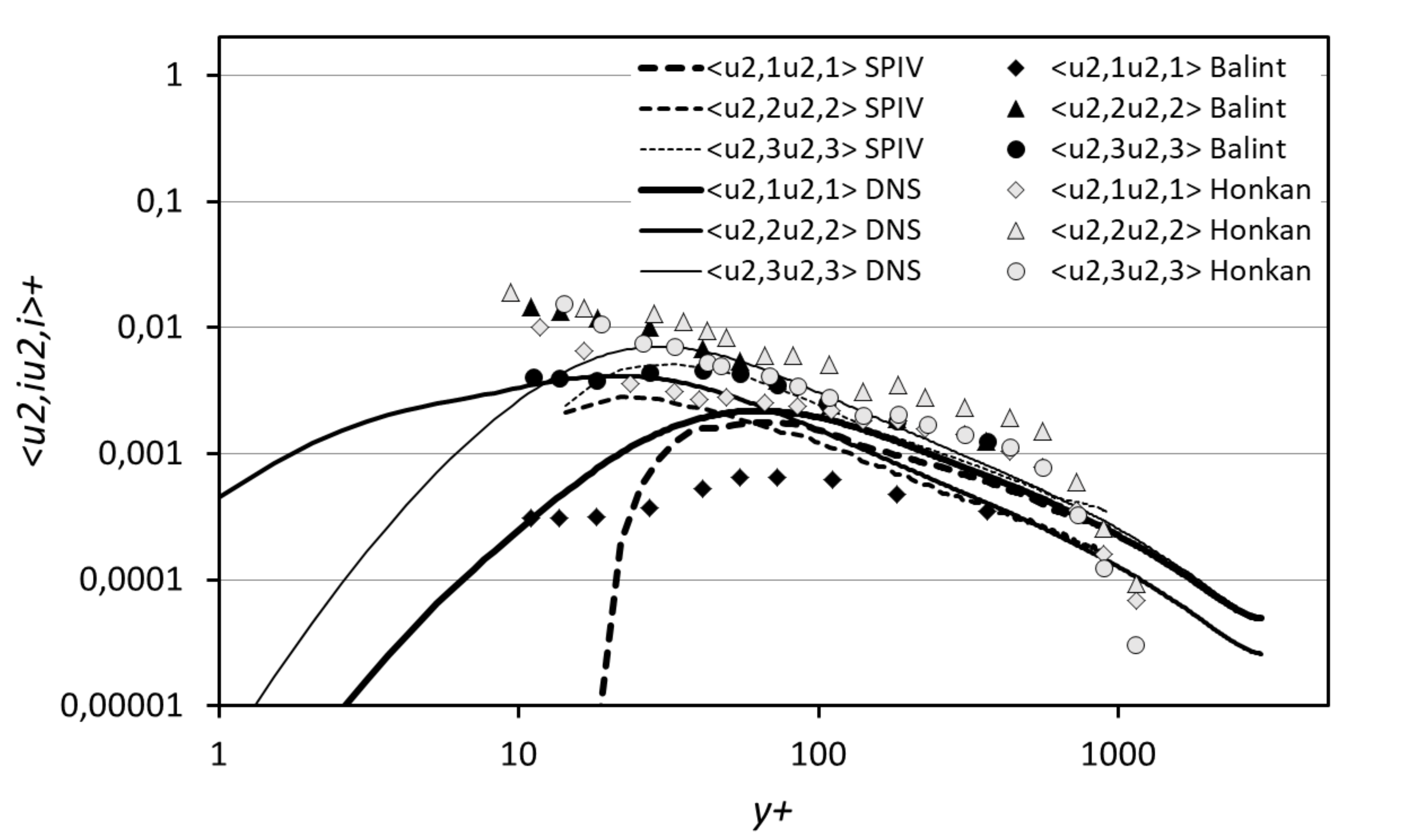}}
	\resizebox{0.76\linewidth}{!}{\includegraphics[scale=1]{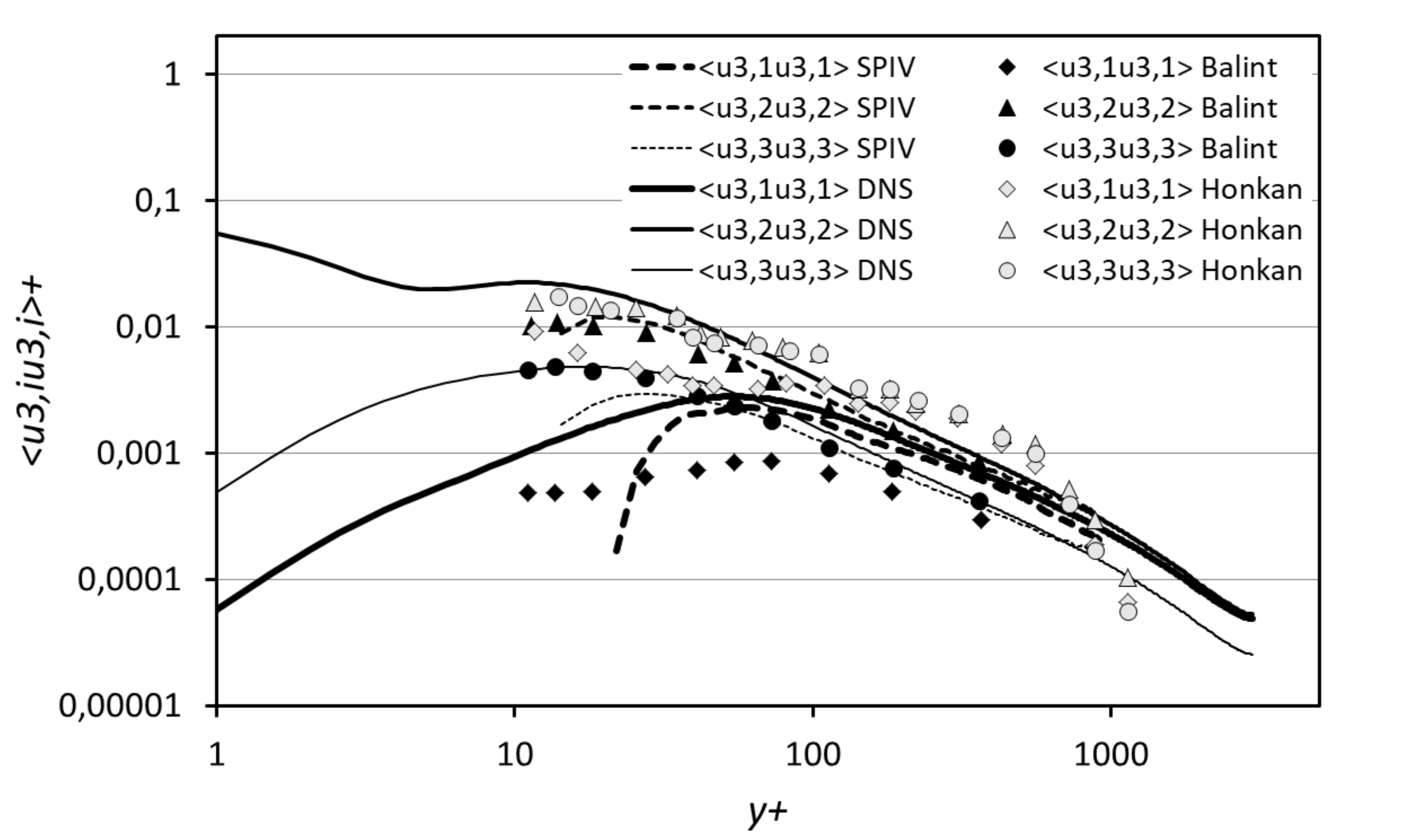}}
	\caption{Comparison of each derivative variance with DNS.}
	\label{fig:17}      
\end{figure*}

\begin{figure*}[ht]
	\resizebox{0.8\linewidth}{!}{\includegraphics[scale=1]{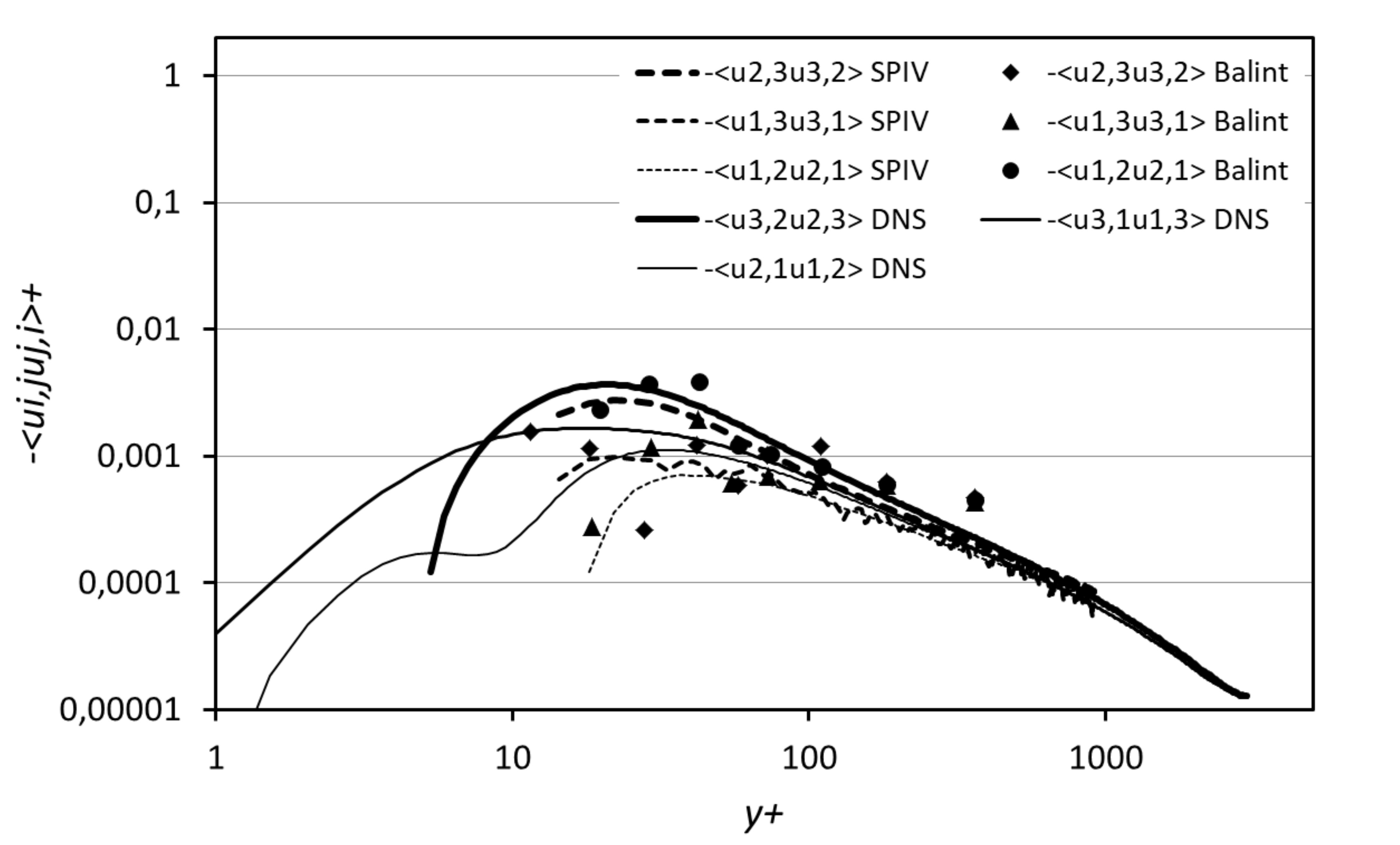}}
	\caption{Comparison of each derivative covariance with DNS.}
	\label{fig:18}      
\end{figure*}

All the derivative moments of equation (\ref{eq:02}) were computed by the method described in section \ref{sec:3}. The term $\left\langle \frac{\partial u_3}{\partial x_1}\frac{\partial u_1}{\partial x_3} \right\rangle$ had to be computed at the intersection of the two SPIV planes and could not be averaged along one of the two planes; so it shows a lower level of convergence. Also plotted in figures \ref{fig:17} and \ref{fig:18} are the results from the DNS of a channel flow of \cite{thais11} at a comparable Reynolds number.  These results are used extensively in parts I and II to analyse the near wall dissipation, so they need not be discussed in detail here. 

\begin{figure*}
	\resizebox{0.85\linewidth}{!}{\includegraphics[scale=1]{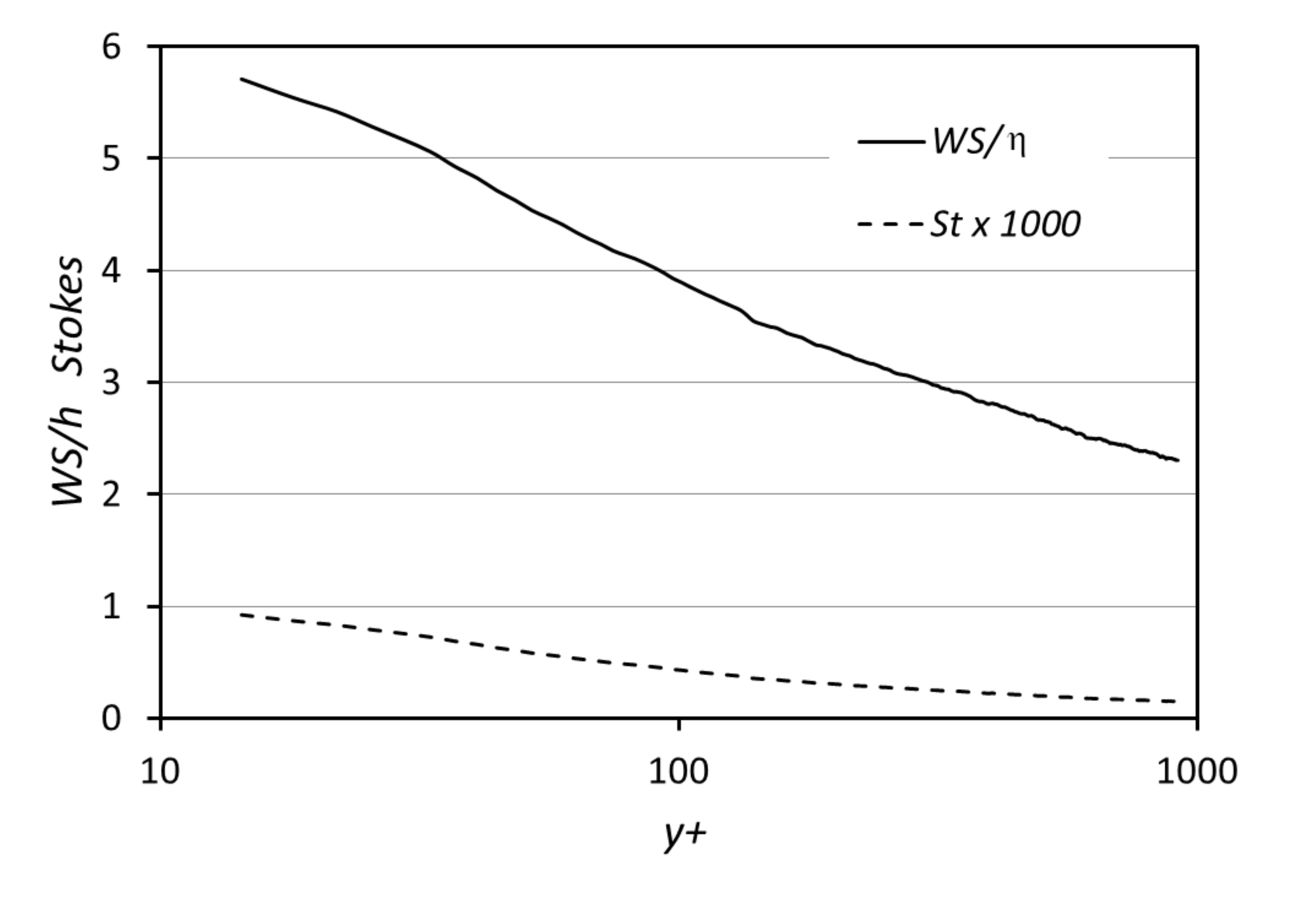}}
	\caption{Interrogation window size normalized by the Kolmogorov micro-scale, $(\nu/\varepsilon)^{1/4}$ and the particle time scale normalized by the Kolmogorov micro-time, $(\nu/\varepsilon)^{1/2}$ as a function of wall distance normalized by $\nu/u_\tau$.}
	\label{fig:19}      
\end{figure*}

Close to the wall the present results clearly underestimates the derivative moments compared to the DNS. This is due to a filtering of the smaller scale as was shown in section \ref{sec:2} and as is illustrated by figure \ref{fig:19} which gives the size of the interrogation window IW compared to the Kolmogorov length scale $\eta$. This gives a clear indication of the level of spatial filtering due to the PIV interrogation window size. This will be revisited in the conclusion to provide some guidelines for performing PIV experiments with an improved assesment of the derivatives moments.

\section{Some new theoretical considerations}

Clearly, measuring the dissipation components in a turbulent flow with PIV is not straightforward. Much care has to be taken both in the set-up of the experiment and in the processing of the PIV data to come to usable results. The challenges presented by the above experiment and the data produced  led many discussions among the authors; and these resulted in the recent theoretical contribution by \cite{georgestanislas20} which revisited the theory of the PIV measurement and of the associated noise. It is not the place here to redevelop this theory, but it is of interest to give the main results, and use them to bring more insight into the present contribution.

Only the results of interest to the present study are included here, all the details can be found in \cite{georgestanislas20} and a summary of the approach in Appendix \ref{appendix}. The first important result relevant for the present analysis is the model of one-dimensional spectrum:

\begin{equation}
E_{1,1PIV}(k_1) = \left({E}_{1,1}(k_1) + \frac{1}{N} \left\{ \langle {u'_1}^2 \rangle  +  \frac{\Delta^2}{12}  \right\}
\frac{\Delta_1}{2\pi} \right)\left[\frac{sin (k_1 \Delta_1/2)}{k_1
	\Delta_1/2}\right]^2 
\label{eq:1Dspectrumplusnoise}
\end{equation} 
where $E_{1,1PIV}(k_1)$ is the measured spectrum, ${E}_{1,1}(k_1)$ is the real volume-averaged spectrum, $N$ is the number of particles in the interrogation volume, $\langle {u'_1}^2 \rangle$ is the streamwise Reynolds stress, $\Delta$ is the quantization of the particle images when digitized by the CCD or CMOS and $\Delta_1$ is the size along $x_1$ of the interrogation volume. This equation is exactly equation (\ref{eq:03s}) with:

\begin{equation}
E_{noise} =  \frac{1}{N} \left\{ \langle {u'_1}^2 \rangle  +  \frac{\Delta^2}{12} \right\} \frac{\Delta_1}{2\pi}
\label{eq:1Dspectrumsnoise}
\end{equation}
As can be seen, $E_{noise}$ is no more a parameter to be adjusted empirically using the data, but one that can be predicted if the turbulence intensity is known.  \cite{georgestanislas20} came to the conclusion that $\Delta = 0.5$ is quite representative of particle images quantization error.

A second interesting result of the theory is the prediction of the noise on the fluctuating velocity moments:

\begin{equation}
\sigma_i^2 =\frac{1}{N} \left\{ \langle u'_i u'_i\rangle + \frac{1}{12}\Delta^2  \right\}W(U_i\delta t) 
\label{eq:rsyerror}
\end{equation}
where $\sigma_i$ is the noise variance on the $i^{th}$ component (no summation on $i$) and $W = \displaystyle\prod_{i = 1}^{3}(1-\frac{U_i\delta t}{\Delta_i})$ is the windowing function which acts only if window shifting is used in the PIV analysis.

The results above can be substituted directly into the noise parameter $\zeta$ (or $E_{noise}$) through the integration of the spectrum as detailed by \cite{foucaut04}:

\begin{equation}
\zeta _{th} = \frac{\sigma ^2\Delta_1\Delta_2}{4 I}
\label{zetasp}
\end{equation}
where $I = 1,492$ is the integral over $[0,2\pi]$ of the ``sinc'' function and $\Delta_1$ and $\Delta_2$ are the dimensions of the interrogation window in $mm$ ($X$ and $Y$ in \cite{foucaut04} notation).

The last result of this theory is that the noise on the derivative moments is twice the noise on the velocity moments:

\begin{equation}
\xi_{ik} =\frac{2}{N} \left\{ \langle u'_i u'_k
\rangle + \frac{1}{12}\Delta^2 \delta_{ik} \right\}W(U_i\delta t) 
\label{eq:differenceerror}
\end{equation}

Based on these equations, and on the parameters of the experiment, it is possible to compare the noise obtained from the above data analysis with the predictions of the theory. This is done in Table \ref{tab:3} for the $x_1,x_2$ configuration given in Table \ref{tab:1}:
\begin{table*}[h]
	\caption{Noise characteristics, comparison between the theory and the data for the $x_1,x_2$ plane}
	\label{tab:3}       
	\begin{center}
	\begin{tabular}{lllllll}
		\hline
		Comp. & IW & Pixels&N&$u_iu_i$&$\Delta$&W   \\
		\hline
		$u_1$&32x24&768&23.04&0.062	 & 0.0007  & 0.75\\
		$u_2$&32x24&768&23.04&0.018  & 0.0007 & 1.00\\
		$u_3$&32x24&768&23.04&0.026	 & 0.0007  & 1.00\\
		\hline
		$\sigma_{SP}$ & $\sigma_{Dif}$ & $\sigma_{th}$&$Ratio_1$&$\zeta$&$\zeta_{th}$& $Ratio_2$  \\
		\hline
		0,020&0.025&0.045&1,83& 1.4E-10  & 6.63E-10&4.73\\
		0,030&0.034&0.028&0,82& 3.08E-10 & 2.57E-10&0.83\\
		0,029&0.034&0.034&0,98& 2.94E-10  & 3.71E-10&1.26\\
		\hline
	\end{tabular}
\end{center}
\end{table*}
where IW is the size of the interrogation window in pixels, Pixels is the number of pixels in the IW, N is the number of particles inside the IW based on $N_p = 0.03$, $u_iu_i$ is the Reynolds stress in $m^2/s^2$ corresponding to the component, $\Delta$ is the quantization error on the particle images in $m$, $W$ is the windowing coefficient, $\sigma_{SP}$ is the noise variance in $m/s$ obtained from the spectrum analysis, $\sigma_{Dif}$ is the noise variance obtained from the difference between the two planes and $\sigma_{th}$ is the theoretical value given by equation (\ref{eq:rsyerror}), $Ratio_1 = \sigma_{th}/\sigma_{Dif}$, $\zeta$ is the noise parameter in SI units deduced from the spectrum analysis, $\zeta_{th}$ is the value provided by equation (\ref{zetasp}) and $Ratio_2 = \zeta_{th}/\zeta$.

As can be seen in this table \ref{tab:3}, the prediction is relatively good along $x_2$ and $x_3$, while the noise level is overestimated along $x_1$. This difference can be explained by the fact that the theory does not take into account the window deformation approach used for the PIV analysis. In the present flow, this technique affects mostly the streamwise component of the Reynolds stress. Taking this point into consideration, the predictions provided in table \ref{zetasp} appear quite satisfactory. Although the theory was developed for 2D2C PIV, it seems to apply also to Stereo PIV if precaution is taken to make the calculations in the physical space (the image space undergoing deformation). 

The main interest of this theoretical approach is that it brings into light the sources of noise in the turbulence PIV data and consequently the way to minimize them. The two main sources are the velocity fluctuations inside the interrogation window and the quantization error when the particle images are digitized by the camera sensor. It appears in \cite{georgestanislas20} and here that the former is generally dominating the later. Looking at equation \ref{eq:rsyerror}, the only parameters on which we can act are the number of particle $N$ in the interrogation window and the windowing coefficient $W$. The second one can affect only the Reynolds stress component in the direction of the mean flow and implies minimizing the overlap between the two shifted windows by adjusting their size and the dynamic range (by adjusting the PIV $\delta t$). The effect of the number of particles is relatively slow (as $1/N$), and is limited by the need to keep the particle images separated from each other. It calls nevertheless for a finer tuning of the particle concentration $N_P$ during the experiment. Finally, if the objective is to access velocity gradients statistics, a joint optimization of the magnification and of the IW size must be looked for.

\section{Summary and Conclusions}
\label{sec:5}

From the experimental point of view, measuring the dissipation is a tremendous challenge. Both because it occurs mostly at small scales, and because it involves only derivatives of the velocity field. Measuring it very near a wall is even harder due to the strong mean velocity gradient involved. The aim of the present contribution was to obtain the best possible assessment of the turbulence dissipation near a wall using SPIV. For that purpose, a joint numerical and experimental approach was used. An SPIV experiment was performed in the LMFL boundary layer wind tunnel. And also the results of a plane channel flow DNS at a comparable Reynolds number were used for comparison and validation. The DNS was especially valuable for providing complementary information to the experiment, especially very near the wall. 

The main problem of PIV is the limited dynamic range of the technique and the relatively high level of noise. This is not so much a problem when looking at the large scales, but it becomes a real challenge when trying to assess the dissipation, due to the shape of the turbulence spectrum which decreases extremely fast toward the high wave numbers. This means that at the dissipative scales the signal-to-noise ratio becomes a critical issue. The present results show that each stage of the experiment and data processing has to be optimized in order to obtain a reasonable estimation of the dissipative terms. 

First of all, the spatial resolution is a key issue. In the present experiment the interrogation window size was varying between roughly 2 and 5 Kolmogorov units from the outer part of the field to the near wall. The results show that some spatial filtering could not be avoided; and if the experiment were to be repeated, staying below 3 Kolmogorov and even close to 2 would be much better. This is quite demanding. In the same conditions, the interrogation window would be close to 0.5 mm in the flow. Luckily, PIV cameras are making continuous progress and this would not be today at the expense of the field-of-view. 

Second of all, the optimisation of the interrogation window size in terms of spectral filtering. The method of \cite{foucaut04}  optimizes the spectral response. It's extension here to the derivative filter minimizes the noise without losing signal. But, to keep this signal, the data are still quite noisy even after these two steps. This is where the cross-planes SPIV set-up and the exploitation of the continuity equation made the difference. These two tools allowed an accurate characterization of the noise on the derivative moments and the removal of this noise, leading to results which were only limited by the spatial resolution. Thanks to this approach, a good agreement was obtained with the DNS and the differences could be explained, leading to a good cross-validation of the two approaches and to an interesting complementarity of the data. 

Besides, thanks to the thinking and exchanges around these data, \cite{georgestanislas20} were able to develop a theory explaining and predicting the noise in PIV measurements. This theory, derived for 2D2C PIV, was applied here nevertheless to the SPIV data, and gave results which are quite comforting in terms of the assessment of the noise level in the data generated by this experiment. This new theory will surely become an interesting tool to help in minimizing the noise in PIV measurements. 

On the basis of the above analyses and results and of the confidence it gave us in our data, parts I \& II of the present contribution (\cite{stanislas20,george20}) allowed us to analyse in detail the dissipation of both the Turbulence Kinetic Energy and the Reynolds stresses in near wall turbulence, and to come to some conclusion about their origin and behaviour near a solid wall. As shown extensively in part I, the best results of the literature  compare quite favourably with the present contribution (\cite{stanislas20}).

\section*{Acknowledgement}
This work was supported through the International Campus on Safety and Inter modality in Transportation (CISIT). This work was carried out within the framework of the CNRS Research Federation on Ground Transports and Mobility, in articulation with the Elsat2020 project supported by the European Community, the French Ministry of Higher Education and Research, the Hauts de France Regional Council. Centrale Lille is acknowledged for providing regular financial support to the subsequent visits of Pr. George. R\'{e}gion Nord Pas de Calais and CNRS are also acknowledged for providing support to these visits. L. Thais is  acknowledged for providing the data of his DNS of channel flow. J.P. Laval is acknowledged for processing the data of L. Thais for the present needs.
\medskip

\noindent{\bf Declaration of Interests. The authors report no conflict of interest}

\appendix

\section{Appendix}
\label{appendix}

As any measuring tool has a limited spatial resolution, the goal of a turbulence experiment is to measure at best the instantaneous
volume-averaged Eulerian velocity, say:

\begin{equation}
\tilde{u}_i(\vec{y},t) =  \frac{1}{V} \int
{u_i}(\vec{x},t) w(\vec{y},\vec{x}-\vec{y}) d\vec{x}
\label{eq:Eulerianinstantvolumeavgdvelocity}
\end{equation}
where $y$ is the position vector of the center of the measuring  volume, $V$ is the volume, $u_i(\vec{x},t)$ is the local Eulerian velocity and the $\tilde{}$ represents volume-averaged quantities. The weighting function $w(\vec{y},\vec{x}-\vec{y})$ defines the measuring volume which, for PIV, can be approximated as a three-dimensional top-hat function.  Note
that all the {\it fluid} particles in the volume
contribute equally to the integral.  Also in the limit, as $V \rightarrow
0$, $w(\vec{y}, \vec{x}-\vec{y})/V \rightarrow \delta(\vec{y})$, so the volume-averaged velocity tends toward the point velocity as the volume
shrinks to a point.

In the case of PIV, we do not have access to this volume-averaged velocity as the recorded information is the image of a limited number of particles (usually about 10) inside an interrogation volume at two instant of time. The idea of \cite{georgestanislas20} was to represent this "PIV" instantaneous velocity  ${u}_{oi}(\vec{y},t)$ as:

\begin{equation}
{u}_{oi}(\vec{y},t) = \frac{1}{N(\vec{y})}\int_{all
	space} {v_i}(\vec{a},t) w(\vec{y},\vec{X}[\vec{a},t]-\vec{y})
g(\vec{a}) d\vec{a} \label{eq:basicunbiased}
\end{equation}
where $N(\vec{y}) =\langle n(\vec{y},t) \rangle = \mu V(\vec{y})$ is the average number of particles in the measurement volume at $y$, $\mu$ is the particle concentration, $V(\vec{y})$ is the volume and ${v_i}(\vec{a},t)$ is the individual particle Lagrangian  velocity, function of the Lagrangian coordinate $a$ and time $t$.  The function $g(\vec{a})$ is a random locator function giving the initial Lagrangian position of the particles present in the measurement volume at time $t$. It can be represented as a finite sum of delta functions.

Two important concepts were introduced by \cite{georgestanislas20}: first the use of generalized functions and second the Lagrangian representation of the particles. Based on  these concepts, it was possible to develop single and two points statistics, Reynolds stresses and spectra and to explicit the noise contributing to the difference between the PIV measurement and the volume averaged quantity.

\bibliographystyle{plainnat}

\bibliography{biblio}

\begin{thebibliography}{24}
\providecommand{\natexlab}[1]{#1}
\providecommand{\url}[1]{\texttt{#1}}
\expandafter\ifx\csname urlstyle\endcsname\relax
  \providecommand{\doi}[1]{doi: #1}\else
  \providecommand{\doi}{doi: \begingroup \urlstyle{rm}\Url}\fi

\bibitem[Adrian(1997)]{adrian97}
R.~J. Adrian.
\newblock Dynamic ranges of velocity and spatial resolution of particle image
  velocimetry.
\newblock \emph{Measurement Science and Technology}, 8\penalty0 (12):\penalty0
  1393--1398, 1997.

\bibitem[Adrian et~al.(2000)Adrian, Meinhart, and Tomkins]{adrian00}
R.J. Adrian, C.D. Meinhart, and C.D. Tomkins.
\newblock Vortex organization in the outer layer of the turbulent boundary
  layer.
\newblock \emph{J. Fluid Mech.}, 422:\penalty0 1--54, 2000.

\bibitem[Carlier and Stanislas(2005)]{carlier05}
J.~Carlier and M.~Stanislas.
\newblock Experimental study of eddy structures in a turbulent boundary layer
  using particle image velocimetry.
\newblock \emph{J. Fluid Mech.}, 535:\penalty0 143--188, 2005.

\bibitem[Coudert and Schon(2001)]{coudert01}
S.~Coudert and J.~P. Schon.
\newblock Back projection algorithm with misalignment corrections for {2D3C}
  {Stereoscopic PIV}.
\newblock \emph{Measurement Science and Technology}, 12:\penalty0 1371--1381,
  2001.

\bibitem[Foucaut and Stanislas(2002)]{foucaut02}
J.~M. Foucaut and M.~Stanislas.
\newblock Some considerations on the accuracy and frequency response of some
  derivative filters applied to {PIV} vector fields.
\newblock \emph{Measurement Science and Technology}, 13:\penalty0 1058--1071,
  2002.

\bibitem[Foucaut et~al.(2004)Foucaut, Carlier, and Stanislas]{foucaut04}
J.~M. Foucaut, J.~Carlier, and M.~Stanislas.
\newblock {PIV} optimization for the study of turbulent flow using spectral
  analysis.
\newblock \emph{Measurement Science and Technology}, 15:\penalty0 1046--1058,
  2004.

\bibitem[Foucaut et~al.(2014)Foucaut, Coudert, C., and C.M.]{foucaut14}
J.~M. Foucaut, S~Coudert, Braud C., and Velte C.M.
\newblock Influence of light sheet separation on spiv measurement in a large
  field spanwise plane.
\newblock \emph{Measurement Science and Technology}, 25\penalty0 (3), 2014.

\bibitem[Foucaut et~al.({2011})Foucaut, Coudert, Stanislas, and
  Delville]{foucaut11}
J.-M. Foucaut, S.~Coudert, M.~Stanislas, and J.~Delville.
\newblock Full 3d correlation tensor computed from double field stereoscopic
  piv in a high reynolds number turbulent boundary layer.
\newblock \emph{Experiments in Fluids}, {50}\penalty0 ({4, Sp. Iss.
  SI}):\penalty0 {839--846}, {APR} {2011}.

\bibitem[Ganapathisubramani et~al.(2005)Ganapathisubramani, Hutchins,
  Hambleton, Longmire, and Marusic]{ganapathisubramani05a}
B.~Ganapathisubramani, N.~Hutchins, W.~T. Hambleton, E.~K. Longmire, and
  I.~Marusic.
\newblock Investigation of large-scale coherence in a turbulent boundary layer
  using two-point correlations.
\newblock \emph{Journal of Fluid Mechanics}, 524:\penalty0 57--80, 2005.

\bibitem[Ganapathisubramani et~al.(2006)Ganapathisubramani, Longmire, and
  Marusic]{ganapathisubramani06}
B.~Ganapathisubramani, E.~K. Longmire, and I.~Marusic.
\newblock Experimental investigation of vortex properties in a turbulent
  boundary layer.
\newblock \emph{Physics of Fluids}, 18\penalty0 (055105.114), 2006.

\bibitem[George and Stanislas(2020)]{georgestanislas20}
W.K. George and M.~Stanislas.
\newblock On the noise in statistics of {PIV} measurements.
\newblock \emph{Submitted to experiments in Fluids}, 2020.

\bibitem[George et~al.(2020)George, Stanislas, Foucaut, Laval, and
  Cuvier]{george20}
W.K. George, M.~Stanislas, J.~M. Foucaut, J.~P. Laval, and C.~Cuvier.
\newblock Velocity derivatives in a high {R}eynolds number {T}urbulent
  {B}oundary {L}ayers. {P}art 2: {S}tatistical {P}roperties.
\newblock \emph{Submitted to Journal of Fluid Mech.}, 2020.

\bibitem[Hambleton et~al.(2006)Hambleton, Hutchins, and Marusic]{hambleton06}
W.~T. Hambleton, N.~Hutchins, and I.~Marusic.
\newblock Simultaneous orthogonal plane particle image velocimetry measurements
  in a turbulent boundary layer.
\newblock \emph{Journal of Fluid Mechanics}, 560:\penalty0 53--64, 2006.

\bibitem[Herpin et~al.(2012)Herpin, Coudert, Foucaut, and Stanislas]{herpin11}
S.~Herpin, S.~Coudert, J.-M. Foucaut, and M.~Stanislas.
\newblock Influence of the reynolds number on the vortical structures in the
  logarithmic region of turbulent boundary layers.
\newblock \emph{Journal of Fluid Mechanics}, 716:\penalty0 5--50, 2012.

\bibitem[Kahler and Stanislas(2000)]{kahler00}
C.~J. Kahler and M.~Stanislas.
\newblock Investigation of wall bounded flows by means of multiple plane stereo
  {PIV}.
\newblock In \emph{to appear in the proceedings of the 10th Int. Symp. on Appl.
  of Laser Tech. to Fluid Mech}, Lisbon, 2000.

\bibitem[Lavoie et~al.(2007)Lavoie, De~Gregorio, Romano, and Antonia]{lavoie07}
G.~Lavoie, P. nd~Avallone, F.~De~Gregorio, G.~P. Romano, and R.~A. Antonia.
\newblock Spatial resolution of piv for the measurement of turbulence.
\newblock \emph{Exp. in Fluids}, 43:\penalty0 39--51, 2007.

\bibitem[Raffel et~al.(1998)Raffel, Willert, and Kompenhans]{raffel98}
M.~Raffel, C.~Willert, and J.~Kompenhans.
\newblock \emph{Particle Image Velocimetry}.
\newblock Springler-Verlag Berlin Heidelberg, 1998.

\bibitem[Scarano({2002})]{scarano02}
F~Scarano.
\newblock {Iterative image deformation methods in PIV}.
\newblock \emph{Measurement Science and Technology}, {13}\penalty0
  ({1}):\penalty0 {R1--R19}, {2002}.

\bibitem[Soloff et~al.(1997)Soloff, Adrian, and Liu]{soloff97}
S.~Soloff, R.~Adrian, and Z.~C. Liu.
\newblock Distortion compensation for generalized {S}tereoscopic {P}article
  {I}mage {V}elocimetry.
\newblock \emph{Meas. Science Tech.}, 8:\penalty0 1441--1454, 1997.

\bibitem[Stanislas et~al.(2020)Stanislas, Foucaut, George, Cuvier, and
  Laval]{stanislas20}
M.~Stanislas, J.~M. Foucaut, W.~George, C.~Cuvier, and J.~P. Laval.
\newblock Velocity derivatives in a high {R}eynolds number {T}urbulent
  {B}oundary {L}ayer. {P}art 1: {D}issipation and {E}nergy {B}alance.
\newblock \emph{submitted to JFM}, 2020.

\bibitem[Stanislas et~al.({2007})Stanislas, Okamoto, Kahler, Westerweel, and
  Scarano]{stanislas07}
M~Stanislas, K~Okamoto, CJ~Kahler, J~Westerweel, and F~Scarano.
\newblock {Main results of the third international PIV challenge}.
\newblock \emph{{EXPERIMENTS IN FLUIDS}}, {45}\penalty0 ({1}):\penalty0
  {27--71}, {AUG} {2007}.
\newblock {Sept.19-20, 2005, Pasadena, USA}.

\bibitem[Thais et~al.(2011)Thais, Tejada-Mart{\'i}nez, Gatski, and
  Mompean]{thais11}
L.~Thais, A.~E. Tejada-Mart{\'i}nez, T.~B. Gatski, and G.~Mompean.
\newblock A massively parallel hybrid scheme for direct numerical simulation of
  turbulent viscoelastic channel flow.
\newblock \emph{Comp. \& Fluid}, 43:\penalty0 134--142, 2011.

\bibitem[Wieneke({2005})]{wieneke05}
B~Wieneke.
\newblock {Stereo-PIV using self-calibration on particle images}.
\newblock \emph{Experiments in Fluids}, {39}\penalty0 ({2}):\penalty0
  {267--280}, {2005}.

\bibitem[Willert(1997)]{willert97}
C.~Willert.
\newblock Stereoscopic digital particle image velocimetry for applications in
  wind tunnel flows.
\newblock \emph{Measurement Science and Technology}, 8:\penalty0 1465--1479,
  1997.

\end{thebibliography}
\newpage

\end{document}